\documentclass[reprint,aps,pra,amsmath,amssymb,superscriptaddress,longbibliography]{revtex4-1}
\usepackage{graphicx}% Include figure files
\usepackage{dcolumn}% Align table columns on decimal point
\usepackage{colortbl}
\usepackage{enumitem}% for itemization in roman numerals
\usepackage{epsfig}

\begin{document}

\title{Cat-state generation and stabilization for a nuclear spin through electric quadrupole interaction}

\author{Ceyhun \surname{Bulutay}}
\email{bulutay@fen.bilkent.edu.tr}
\affiliation{Department of Physics, Bilkent University, Ankara 06800, Turkey}

\date{\today}

\begin{abstract}
Spin cat states are superpositions of two or more coherent spin states (CSSs) that are distinctly separated 
over the Bloch sphere. Additionally, the nuclei with angular momenta greater than 1/2 possess a quadrupolar 
charge distribution. At the intersection of these two phenomena, we devise a simple scheme for generating
various types of nuclear spin cat states. The native biaxial electric quadrupole interaction that is readily 
available in strained solid-state systems plays a key role here. However, the fact that built-in strain cannot be switched 
off poses a challenge for the stabilization of target cat states once they are prepared. We remedy this by abruptly
diverting via a single rotation pulse the state evolution to the neighborhood of the fixed points of the underlying 
classical Hamiltonian flow. Optimal process parameters are obtained as a function of electric field gradient 
biaxiality and nuclear spin angular momentum. The overall procedure is seen to be robust under 5\% deviations 
from optimal values. We show that higher level cat states with four superposed CSS can also be formed using three 
rotation pulses. Finally, for open systems subject to decoherence we extract the scaling of cat state fidelity damping 
with respect to the spin quantum number. This reveals rates greater than the dephasing of individual
CSSs. Yet, our results affirm that these cat states can preserve their fidelities for 
practically useful durations under the currently attainable decoherence levels.
\end{abstract}

% PhySH
% Research Areas / Optics & lasers / Quantum state engineering
% Research Areas / Quantum information processing / Quantum information processing with continuous variables
% Techniques / Experimental Techniques / Resonance techniques / Nuclear quadrupole resonance
% Physical Systems / Quantum spin models

\maketitle

\section{Introduction}
% Nuclear spin systems and single nuclear spin detection
In the midst of the so-called second quantum revolution \cite{dowling03}, nuclear spin systems have been among the first  
to be proposed and tested \cite{cory97,kane98}.
In due course, the overwhelming majority of implementations have utilized an {\em ensemble} of nuclear spins which stems from
the established bulk nuclear magnetic resonance-based manipulation and detection schemes \cite{levitt07}. 
For quantum information processing, working with a {\em single} spin is desirable to alleviate issues arising from ensemble 
averaging, however, it was initially hindered by the poor detectability \cite{warren97}. More than two decades ago the single 
electron spin detection within a host crystal was achieved \cite{kohler93,wrachtrup93}. 
In the case of a single nuclear spin the remaining challenge was its about two thousand times smaller magnetic moment 
compared to electron \cite{levitt07}. The breakthrough came with the aid of optically detected electron nuclear double 
resonance \cite{wrachtrup97}. Subsequently, the optical readout of a single nuclear spin in a nitrogen-vacancy (NV) 
defect center in diamond was announced \cite{jelezko04b,dutt07}. The next milestone reached on this front was the single-shot 
readout of a single nuclear spin, again within the NV system at room temperature \cite{neumann10,robledo11,dreau13}. 
As other solid-state examples, and using an electrical readout scheme, the implementations on a Tb nuclear spin 
of a single-molecule magnet \cite{vincent12}, and a $^{31}$P donor nuclear spin in silicon \cite{pla13} can be 
mentioned; for a very recent review, see Ref.~\cite{suter17}.
 
% Quantum processing with spins
Such a control on the single spin level in a solid-state system opens enormous opportunities for quantum 
technologies \cite{dowling03}.
Mainly because spin offers an excellent framework for demonstrating interesting quantum states and effects such as 
coherent states \cite{radcliffe71}, squeezing \cite{kitagawa93,ma11}, to name but a few. An important class in quantum 
mechanics is the cat state which corresponds to macroscopically-separated coherent superpositions of coherent states 
\cite{dodonov74,yurke86}. Therefore, its realization in various spin systems has been the aspiration for 
a number of proposals lately, such as the generation of spin cat states in Rydberg atoms \cite{opatrny12}, in Bose-Einstein 
condensates \cite{lau14}, in a finite 
Kerr medium for the big spin-qubit \cite{dooley13} or spin star model \cite{dooley14}. In terms of applications,
spin cat states are suggested for high-precision measurements via dissipative quantum systems of Bose atoms
\cite{huang15}; an exhaustive review of quantum metrology is available from Ref.~\cite{pezze16}.
Unfortunately, these are either of model level \cite{dooley13,dooley14}, or based on atomic nonlinearities \cite{pezze16}, 
arising from Rydberg blockade \cite{opatrny12}, or collisional effects in Bose condensed atoms \cite{lau14,huang15}. 
Therefore, for the case of a nuclear spin in a solid-state environment these recipes are of no avail.

% This work
In this work our aim is to present a simple means to generate different kinds of cat states on a single nuclear spin by 
harnessing the quadrupole interaction (QI) \cite{cohen57,das58} which intrinsically operates on the quadrupolar 
nuclei, that is, with spin quantum number greater than 1/2. Motivated by a recent experimental exposition 
of squeezing with spin-7/2 nuclei \cite{auccaise15}, this work builds upon our prior study where we have shown 
that the generic QI supports continuous tuning of squeezing from one-axis to two-axis countertwisting limits \cite{korkmaz16}.
We confine ourselves to the spin values between 1 and 9/2 that correspond to the range of abundant isotopes of
quadrupolar nuclei. We adopt new concepts developed for freezing the spin squeezing \cite{jin07,wu15,kajtoch16}, 
in our case to stabilize the cat states once they are produced. The robustness of our scheme is 
checked under various drifts or errors in the process parameters \cite{cummins03}. Furthermore, going up to 
the next level in hierarchy, we consider the production of the superpositions of spin cat states 
that is essential for the quantum error correction against spin flips without revealing the registered quantum 
information \cite{mirrahimi14}. The kind that we discuss corresponds to a rotating spin cat state superposed 
to a fixed counterpart, enabling a relative phase accumulation, that is composed of the so-called moving coherent 
states \cite{albert16}. We assure the longevity of these states by exhibiting their resilience to phase decoherence 
under realistic conditions.

% Plan
The paper is organized as follows. In Sec.~II we present the background theoretical information as well as our notation on QI, 
coherent spin and cat states, measures used for assessment, decoherence model, and the phase portraits of the 
QI Hamiltonian. In Sec.~III we report our results starting with the overall operation, followed by optimization of 
performance, and its sensitivity analysis; we then discuss the extension of cat-state generation to their superpositions,
and how decoherence affects the cat states. We also comment on practical aspects, and potential applications.
Our main findings are summarized and suggestions for future directions are 
outlined in Sec.~IV. The Appendix section contains the derivation of the closed-form expression for the time evolution operator 
under a certain case that is required in the main text.

\section{Theory}
\subsection{Quadrupole spin Hamiltonian}
Nuclei with spin $I\ge 1$ are named as quadrupolar because of their multipolar charge distributions. This makes them 
susceptible to electric field 
gradients (EFG) \cite{cohen57,das58}. The latter is routinely present within a solid-state matrix, predominantly being caused 
by strain \cite{bulutay12}. The EFG at a nuclear spin site is described by a second-rank tensor
involving second-order spatial derivatives of the crystal potential $V$,
$
V_{ij} \equiv \partial^2V/\partial x_i\partial  x_j \, ,
$ 
which becomes diagonal for a particular orientation of coordinate axes. 
This is referred to as the principal EFG axes which is what we shall use throughout this work. 
Here, as a convention, the Cartesian axes are labeled in the way that the nonvanishing EFG components obey the ordering 
$|V_{zz}|\geq |V_{yy}|\geq |V_{xx}|$. 

The EFG acts on the spin degrees of freedom of a quadrupolar nucleus
as governed by the Hamiltonian \cite{cohen57}
\begin{equation}
\label{H-QI}
\hat{H}_\eta = \frac{hf_Q}{6}\left[ 3\hat{I}_z^2 - \hat{I}^2 + \eta \left(\hat{I}_x^2-\hat{I}_y^2\right) \right]\, ,
\end{equation}
where, $h$ is Planck's constant, and $f_Q$ is the quadrupole linear frequency controlled by the EFG major principal value, $V_{zz}$.
Since we shall not consider any other steady term in the Hamiltonian, $f_Q$ will serve for setting the time scale of the 
dynamics; typical values will be stated when we discuss decoherence processes. The magnitude of spin angular momentum 
vector $I$ is conserved, and as in our prior work \cite{korkmaz16}, this term can be dropped from dynamics at will.
Note that we parametrize the Hamiltonian with respect to $\eta=\left(V_{xx}-V_{yy}\right)/V_{zz}$ which
defines the degree of biaxiality  of the EFG \cite{cohen57,das58}, and as we shall see, plays a central 
role for the cat-state generation 
and stabilization. It is confined to the range $[0,1]$: the lower limit corresponds to a cylindrically symmetric 
EFG distribution, while the upper limit $\eta=1$ can be realized, for instance, in two-dimensional materials \cite{korkmaz16}.

\subsection{Coherent spin and cat states}
A coherent spin state (CSS) centered around the spherical angles $(\theta,\phi)$ can be obtained from the $z$-oriented Dicke 
spin state $\left|I, I_z=I\right\rangle$ via
\begin{equation}
\left|\theta,\phi\right\rangle = \exp\left[ i\theta\left(\sin\phi\,\hat{I}_x-\cos\phi\,\hat{I}_y\right)\right] \left|I, I\right\rangle\, ,
\end{equation}
where the operator on the right performs rotation around the axial vector $(\sin\phi,-\cos\phi, 0)$ by an angle $\theta$ 
\cite{sanders89}. For convenience we shall denote the CSS located around the six axial Cartesian directions over the Bloch sphere as 
$\left|\pm X\right\rangle$, $\left|\pm Y\right\rangle$, and $\left|\pm Z\right\rangle$.
We emphasize that the Cartesian directions here are not arbitrary, but are based on EFG principal axes.
Once again, in relation to the Dicke state, $|j,m\rangle$ as labeled with total angular momentum $j$, and its quantization-axis 
projection $m$, we have for the $z$- quantization axis $| \pm Z\rangle=| I, \pm I\rangle$, while 
$| \pm X\rangle$ and $| \pm Y\rangle$ are their rotated forms around the $y$ and $x$ axes, respectively.

For quantum metrology and other quantum information technologies, it is the coherent superpositions of such CSSs that are of
importance, especially, if they correspond to macroscopically-distinguishable superpositions, so-called cat states 
\cite{gilchrist04}. Historically, the even and odd cat states were first to be introduced, having the forms
$\mathcal{N} \left[\left| \alpha\right\rangle + \left| -\alpha\right\rangle\right]$ and
$\mathcal{N} \left[\left| \alpha\right\rangle - \left| -\alpha\right\rangle\right]$ 
respectively \cite{dodonov74}, where $\left| \alpha\right\rangle$ denotes a generic coherent state \cite{gerry-knight}.
This is followed by the so-called Yurke-Stoler state, $\mathcal{N} \left[\left| \alpha\right\rangle \pm i \left| -\alpha\right\rangle\right]$
\cite{yurke86}. In these expressions, $\mathcal{N}$ is the normalization factor, which is different for each case; 
in the remainder of this work, for brevity we drop them from our subsequent notations. 
Additionally, in our following discussion, we prefer the terms, 
equator-bound and polar-bound target cat states, for
$\left[\left| Y\right\rangle +e^{i\varphi} \left| -Y\right\rangle\right]$, and
$\left[\left| Z\right\rangle +e^{i\varphi} \left| -Z\right\rangle\right]$
respectively, according to where the cat pair is located on the Bloch sphere.
The remaining $x$ axis, paired with the minor EFG component, will serve as the main rotation axis.
Such coherent superpositions of maximally separated two CSS over the Bloch sphere will be termed as 
$N=2$ cat states. In the Results section we shall also introduce the coherent superpositions of two maximally-separated 
$N=2$ cat states making up a $N=4$ cat state.

\subsection{Measures}
To quantify how closely a generated state $\left| \psi\right\rangle$ reproduces a certain target state 
$\left| \beta\right\rangle$, a common measure is the fidelity which for such pure states becomes simply $F=\left| \langle\beta | 
\psi\rangle\right|$ \cite{nielsen-book}. 
Alternatively, rather than comparing with a {\em fixed} target state, one can use 
absolute macroscopicity measures \cite{bjork04}. These are generally based on the quantum Fisher information in regard 
to an operator/measurement $\hat{A}$, which reduces for a pure state $\left| \psi\right\rangle$ (as invariably considered in this work)
to $\mathcal{F}(\psi,\hat{A})= 4\, \mathcal{V}_\psi(\hat{A})$,
where, $\mathcal{V}_\psi(\hat{A})=\langle\psi|\hat{A}^2|\psi\rangle-\langle\psi|\hat{A}|\psi\rangle^2$ is the variance.
For a spin-$I$ system the effective size is then defined as 
\begin{equation}
\label{eff-QFI}
N_{\rm eff}^{\rm F}(\psi)= \max_{\hat{A} \in \mathcal{A}} \mathcal{F}(\psi,\hat{A})/(2 I)\, ,
\end{equation}
where one maximizes over operators within the relevant set $\mathcal{A}$. To quantify the degree of {\em catness} 
of a superposed state $|\psi_S\rangle=(|\psi_a\rangle+|\psi_b\rangle)/\sqrt{2}$, 
the relative quantum Fisher information (rQFI) has been proposed \cite{frowis12} as 
\begin{equation}
N_{\rm eff}^{\rm rF}(\psi_S)= \frac{N_{\rm eff}^{\rm F}(\psi_S)}{\left[N_{\rm eff}^{\rm F}(\psi_a)+N_{\rm eff}^{\rm F}(\psi_b)\right]/2} \, .
\end{equation}
For pure states, and choosing as the relevant interferometric measurement operators the spin along direction $u$ (i.e., $\hat{I}_u$), each 
maximization in Eq.~(\ref{eff-QFI}) becomes trivial, yielding
\begin{equation}
N_{\rm eff}^{\rm rF}(\psi_S)= \frac{2\, \mathcal{V}_S(\hat{I}_S)}{\left[ \mathcal{V}_a(\hat{I}_a)+\mathcal{V}_b(\hat{I}_b)\right]} \, .
\end{equation}
The variance for a CSS is simply $I/2$, and in the case of a diametrically opposite cat state, for instance, choosing 
one of the target states 
mentioned above, $|\psi_S\rangle\to\left[\left| Z\right\rangle +e^{i\varphi} \left| -Z\right\rangle\right]$, we have 
$\hat{I}_S\to\hat{I}_z$ which yields $\mathcal{V}_S(\hat{I}_S)=I^2$; 
therefore, the maximum value of $N_{\rm eff}^{\rm rF}$ becomes $2I$, which indeed corresponds to largest 
possible separation over the Bloch sphere, namely, its diameter. Thus, to quantify the catness of an evolving state $\psi$, 
for a spin measurement along a direction $u$ we use {\em normalized} rQFI, as
\begin{equation}
\label{rQFI}
\overline{N}_{\rm eff}^{\rm rF}(\psi)= \frac{\mathcal{V}_\psi(\hat{I}_u)}{I^2} \, ,
\end{equation}
which ranges between 0 and 1.

\subsection{Accounting for decoherence}
There is a well-defined phase relation among the constituent CSSs reflecting the coherence of the superposition. 
As such, they are particularly vulnerable to phase noise. The resultant decoherence can be tracked via the system 
density operator using the Lindblad master equation \cite{nielsen-book}
\begin{eqnarray}
\label{lindblad-eq}
\frac{\mathrm{d}}{\mathrm{d}t} \hat{\rho}_S(t) & = & -\frac{i}{\hbar}\left[ \hat{H}, \hat{\rho}_S(t) \right] \nonumber \\
& & + \sum_{m=1}^{2I} \left[\hat{L}_m\hat{\rho}_S(t)\hat{L}^\dagger_m
-\frac{1}{2}\left\{\hat{L}^\dagger_m\hat{L}_m,\hat{\rho}_S(t)\right\}\right] \, ,
\end{eqnarray}
where $\hat{\rho}_S$ is the nuclear-spin density operator, and $[\, ,]$ and $\{\, ,\}$ represent commutator and anticommutator, respectively.
$\hat{L}_m$ is a so-called Lindblad operator characterizing the nuclear spin's coupling to its environment \cite{nielsen-book}. 
For the phase-flip channel of a spin-$I$ system they can be extracted from the associated Kraus operators \cite{pirandola08} as
\begin{equation}
\label{lindblad-op} 
\hat{L}_m=\sqrt{\frac{(2I)!}{m!(2I-m)!} \left(\frac{1-e^{-\gamma}}{2}\right)^m \left(\frac{1+e^{-\gamma}}{2}\right)^{2I-m}} \hat{I}_z^m\, ,
\end{equation}
where $\gamma=1/T_2$ is the dephasing rate, with the well-known coherence dephasing time constant being $T_2$, which is routinely 
measured with spin-echo techniques \cite{levitt07}. 
Even though we shall be using this full Lindblad set, it can be readily verified that in 
the weak damping limit the Lindblad operators reduce to a single one $\sqrt{\gamma I} \hat{I}_z$, as considered in Ref.~\cite{korkmaz16}.

\subsection{Fixed points and their biaxiality dependence}
In the case of spin squeezing, the maximally squeezed quadrature is attained only for an instant over each (quasi-)period 
of the cycle \cite{korkmaz16}. To break away from this regime with large swings Kajtoch 
{\em et al.} proposed to apply a rotation operation when maximum squeezing is reached, and transfer the subsequent flow to regions 
around the fixed points of the classical Hamiltonian, where oscillation amplitudes can be highly suppressed \cite{kajtoch16}.
To implement this recipe for the QI under an arbitrary biaxiality $\eta$,  we need the associated fixed points \cite{huang12}.

\begin{figure}
\includegraphics[width=\columnwidth]{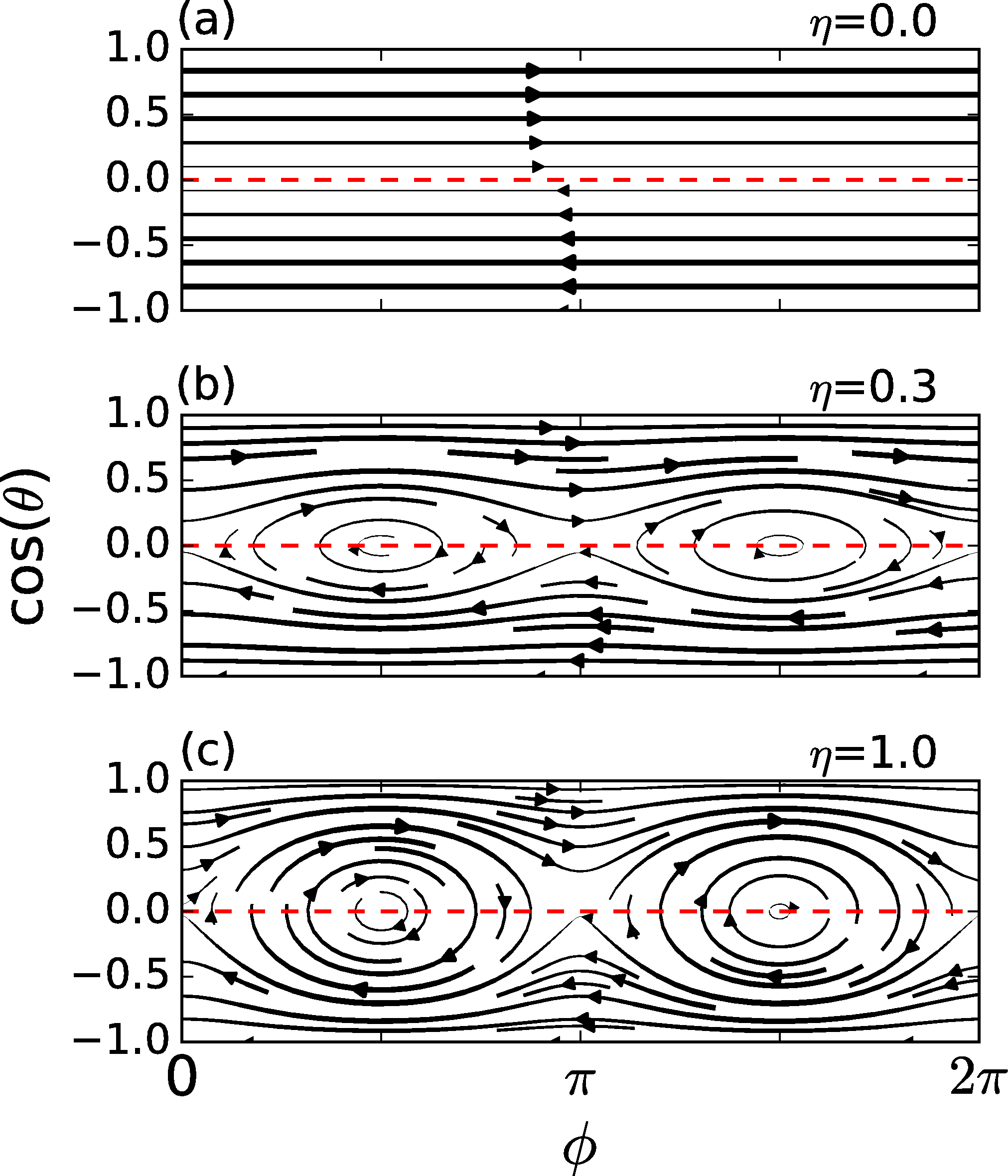}
\caption{Phase portraits obtained from Eqs.~(\ref{Eq-portrait-1}) and (\ref{Eq-portrait-2}) for three different $\eta$
values. Dashed red lines mark the equator on the Bloch sphere. The thickness of the lines is proportional to the speed of the flows.
}
\label{fig1}
\end{figure} 

For a classical spin vector pointing toward the $(\theta, \phi)$ direction, the QI Hamiltonian in Eq.~(\ref{H-QI}) takes the form
\begin{equation}
H_\eta(\theta, \phi) = \frac{hf_QI}{6}\left[ 3\cos^2\theta + \eta\sin^2\theta \cos 2\phi\right]\, ,
\end{equation}
through which the Hamilton equations of motion are obtained for the canonically conjugate variables 
$(\phi, P_\phi\equiv\cos\theta)$ as
\begin{eqnarray}
\label{Eq-portrait-1}
\dot{\phi} & = \frac{hf_QI}{3} P_\phi \left( 3-\eta \cos 2\phi\right) \, , \\
\label{Eq-portrait-2}
\dot{P_\phi} & = \frac{hf_QI}{3} \eta \left( 1-P^2_\phi \right)\sin 2\phi \, .
\end{eqnarray}
The corresponding phase portraits are shown in Fig.~\ref{fig1} for three different $\eta$ values.
Two stable center fixed points lie at the poles ($\theta=0,\pi$) for any $\eta$. Additionally, 
for the case of $\eta=0$ the whole equator line ($\theta=\pi/2$) turns into fixed ``points'', 
whereas for $\eta\ne 0$ they reduce solely to four points at $\pm x$- and $\pm y$-axes i.e., 
$\phi = \{ 0,  \pi, \pi/2, 3\pi/2 \}$. In the latter 
case, $\phi = \{ 0,  \pi\}$ are unstable, and those at $\phi = \{\pi/2, 3\pi/2\}$ are of stable center type fixed points.
In other words, for non-zero $\eta$ the stable fixed points over the Bloch sphere are positioned at  $\pm y$- and $\pm z$-axes 
that we term as equator- and polar-bound, respectively.

\begin{figure*}
\includegraphics[width=1.7\columnwidth]{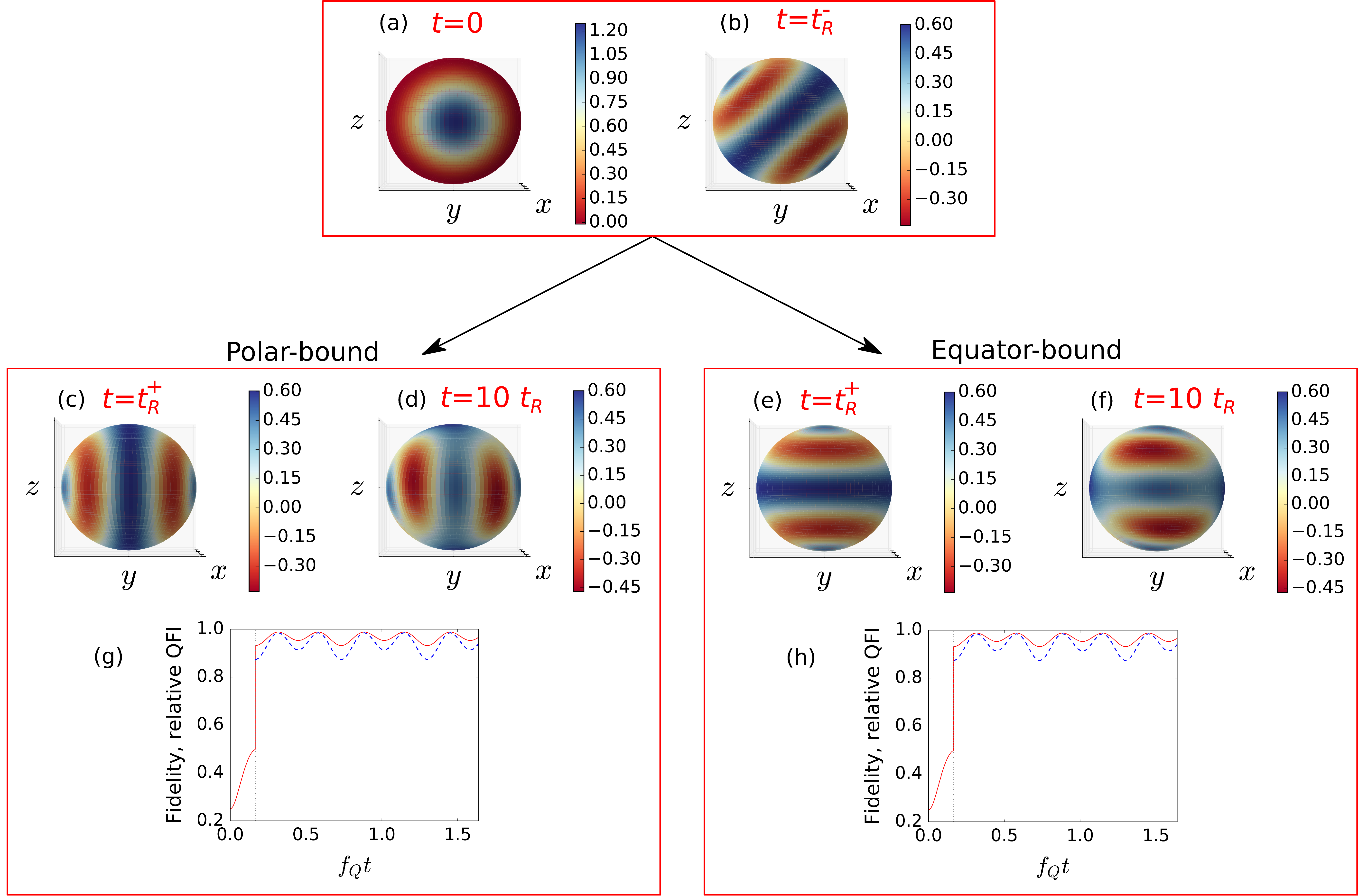}
\caption{The spin Wigner quasi-probability distributions at the main stages of the 
procedure starting from (a) an initial CSS, (b) just before the rotation instant $t=t_R^-$, (c)+(e) just after rotation 
to pole and equator planes $t=t_R^+$, (d)+(f) much later at $t=10 t_R$. Items (g) and (h) show the fidelity (solid/red) 
and normalized rQFI (see, Eq.~(\ref{rQFI})) (dashed/blue) for the polar- and equator-bound evolutions. 
A 5/2-spin with $\eta=1$ is considered. Vertical dotted lines mark the instances of the rotation pulses.
}
\label{fig2}
\end{figure*}

\section{Results}
\subsection{Basic operation}
% Wigner QPD Evolutions
To demonstrate the overall procedure, we consider a 5/2-spin with an initial CSS lying on the $+x$-axis of the 
Bloch sphere, i.e., $|+X\rangle\equiv\left|\theta_{CSS}=\pi/2, \phi_{CSS}=0\right\rangle$ (Fig.~\ref{fig2}(a)). 
Under the action of $\hat{H}_\eta$ (for this example, $\eta=1$), it first goes through a squeezing stage with the antisqueezed axis 
having rotated by about $\pi/4$ from the equatorial plane over the Bloch sphere (Fig.~\ref{fig2}(b)). We terminate this regime suddenly by applying 
a rotation around the $+x$-axis that aligns the spin distribution elongation toward either polar (Fig.~\ref{fig2}(c)) or equatorial 
plane (Fig.~\ref{fig2}(e)), coinciding with the two fixed points of the QI Hamiltonian, as discussed in the previous section. 
Hence, further evolution of the dynamics gets 
localized around the two antipodal fixed points, giving rise to either polar-bound ($\left|\pm Z\right\rangle$; 
see, Fig.~\ref{fig2}(d)) or equator-bound ($\left|\pm Y\right\rangle$; see, Fig.~\ref{fig2}(e)) cat states. 
For all cases, we resort to spin Wigner quasi-probability distribution plots \cite{qutip1,qutip2} which is sensitive 
to phases \cite{gerry-knight}. Observe that in between these antipodal regions, additional fringes, hallmark of quantum 
coherence exist \cite{agarwal97}, as colloquially referred to as the smile of the cat \cite{chumakov99}. 

The degree of success is quantified with the fidelity (Figs.~\ref{fig2}(g) and (h)) reaching typically a maximum value around 
0.95 and a ripple of about 0.05; see, Eqs.~(\ref{Fmax}) and (\ref{Fripple}) below. The normalized rQFI 
calculated by Eq.~(\ref{rQFI}) (shown by dashed lines) follows the same behavior of the fidelity, but with a larger ripple. 
As the rQFI measure simply tracks the separation of the constituent cats, and is not anchored to a target (unlike fidelity),
the valuable conclusion this provides is that concerted deviation from unity under both measures cannot originate from 
simply a rigid oscillation around the target state.

\begin{figure}
 \includegraphics[width=\columnwidth]{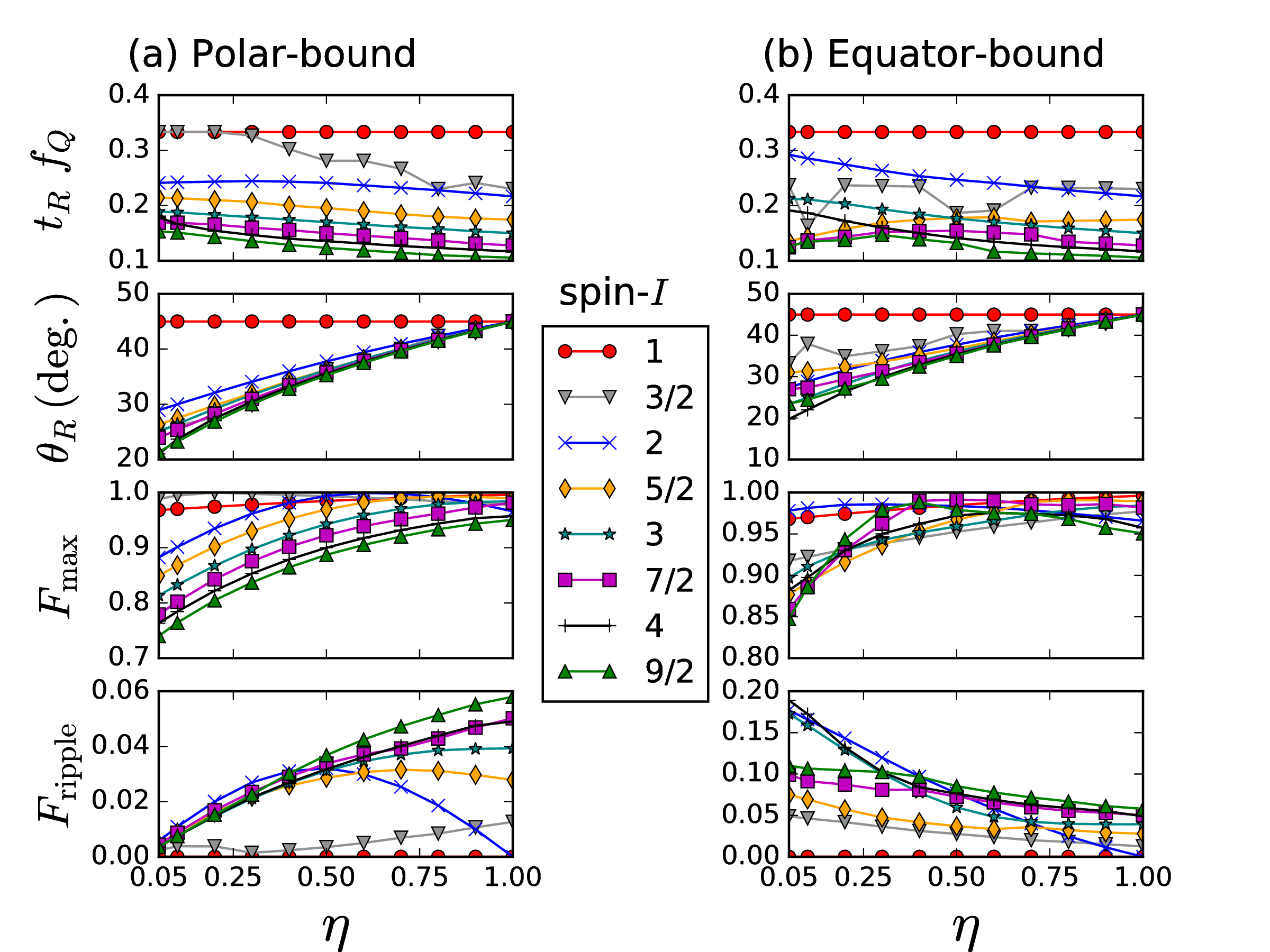}
\caption{Rotation instants and angles (in degrees) for the optimal cat-state generation and stabilization as
quantified by maximum fidelity $F_{\rm max}$ (Eq.~(\ref{Fmax})), and its ripple $F_{\rm ripple}$ (Eq.~(\ref{Fripple}))
for polar- and equator-bound cases.
}
\label{fig3}
\end{figure}

\subsection{Search for optimality}
For each quadrupolar spin from $I=1$ to 9/2, we optimize with respect to the time instant, $t_R$ and the angle 
of the rotation, $\theta_R$
that will orient the major axis of the spin distribution toward the fixed points on either the poles ($\pm z$-axes), or 
equator ($\pm y$-axes). We also let the phase angle, $\varphi$ between the constituent CSSs of the target cats
$\left[ |Z\rangle +e^{i\varphi}|-Z\rangle\right]$ and $\left[ |Y\rangle +e^{i\varphi}|-Y\rangle\right]$,
to be yet another optimization parameter.
Our two optimality criteria are 
\begin{eqnarray}
\label{Fmax}
F_{\rm max} & = & \max F \, , \\
\label{Fripple}
F_{\rm ripple} & = & (\max F - \min F)/2 \, ,
\end{eqnarray}
that is, high fidelity with the associated target cat states and low ripple around the mean fidelity once 
in the stabilization stage, i.e., after the rotation instant. We assign the relative weights 0.55 and 0.45 to these 
two goals, respectively, and obtain corresponding optimal cat-state generation and stability performances.

For the polar-bound targets, this procedure culminates with strictly even cat states, whereas the equator-bound ones exhibit a spin-$I$ 
dependent phase angle, $\varphi=\pi I$, that is, for integer spin nuclei the cat states produced are of the same parity with $I$, and for all 
half-integer spins Yurke-Stoler-type cat states are generated. Figure~\ref{fig3} displays these results as a function of the QI biaxiality parameter, 
$\eta$ the variation of maximum fidelity (Eq.~(\ref{Fmax})), its ripple (Eq.~(\ref{Fripple})), the instants of optimal rotation 
pulse, and the angles around the $+x$-axis required to orient them to the appropriate target planes.
For both polar- and equator-bound cases, fidelity drastically drops when the uniaxiality of QI increases, i.e., $\eta\rightarrow 0$, however,
for the former the ripple in the fidelity also decreases.
This correlates with the fact that the equatorial fixed points soften as $\eta\rightarrow 0$.
In the opposite limit of $\eta\rightarrow 1$ which corresponds to two-axis countertwisting \cite{korkmaz16}, 
the optimal rotation angle goes to $\pi/4$ for all cases, in accordance with the findings of Kajtoch {\em et al.} \cite{kajtoch16}. 
In general, as the spin-$I$ value increases the maximum fidelity reduces.
In this regard, the $I=1$ case appears to show the best performance. However, $I=1$ is actually an outlier with respect to the higher spins, 
showing no dependence to $\eta$ at all. This three-level system also has a similar peculiarity in spin squeezing with exact vanishing of 
uncertainty in one of the quadratures \cite{korkmaz16}. It remains to be seen whether these seemingly impressive $I=1$ performances can be
of any practical relevance.

\begin{figure}
 \includegraphics[width=\columnwidth]{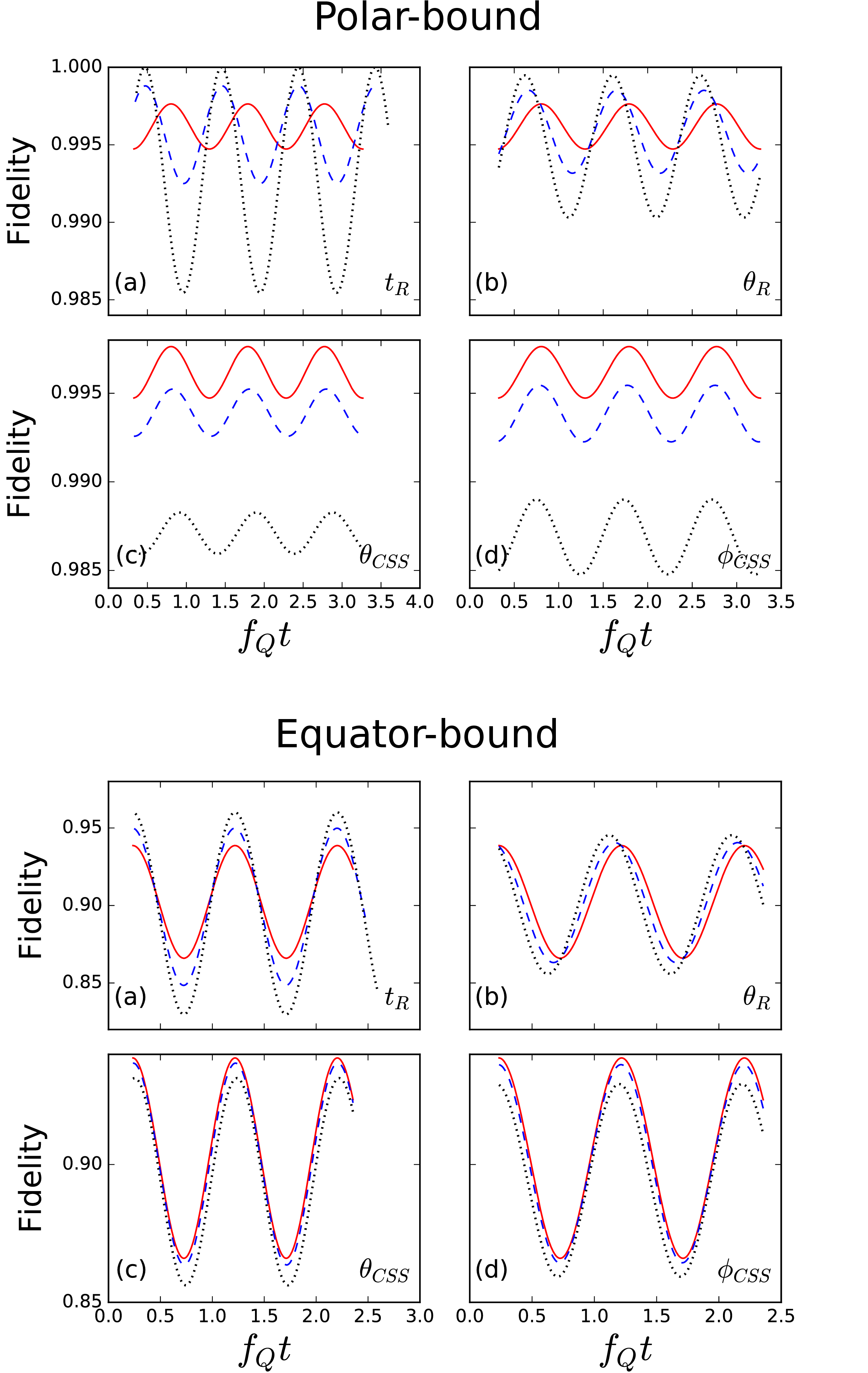}
\caption{For polar- and equator-bound cases, sensitivity of optimal fidelities (solid/red) under 5\% (dashed/blue) and 10\% 
(dotted/black) deviation in the parameters of (a) rotation instant, (b) rotation angle, (c) polar and (d) azimuthal 
offsets in the initial CSS. $I=3/2$, and $\eta=0.3$ is considered.
}
\label{fig4}
\end{figure}

\subsection{Sensitivity}
If the attained optimal conditions in Fig.~\ref{fig3} are only achievable for a very narrow range of parameters, 
it will hamper the
practical utility of the proposed cat-state generation and storage. Therefore, in this section we present the sensitivity analysis 
around the operating points. For this purpose, we choose $I=3/2$ and $\eta=0.3$, where both a very high fidelity and low ripple
values were observed, especially for the polar-bound case (Fig.~\ref{fig3}). The instant when the rotation pulse is applied, $t_R$ 
and the amount of rotation angle $\theta_R$ are the two main parameters here. Additionally, we consider unintentional displacements from 
the assumed location of the initial CSS, $|X\rangle$ along the polar $\theta_{CSS}$ and the azimuthal $\phi_{CSS}$ angles, 
as would be caused when the quadrupolar principal axes are not properly aligned with the CSS or rotation axes. 
This can be termed as the preparation error, encountered, for instance, in the NV centers \cite{robledo11}.
In Fig.~\ref{fig4} we compare the time-dependent fidelity
of the optimal case with those under 5\% and 10\% deviations in each of these parameters. In general terms, a change in 
$t_R$ predominantly increases the ripple around the same mean fidelity value, and this parameter shows higher sensitivity than $\theta_R$
in the same range. $\theta_{CSS}$ and $\phi_{CSS}$ offsets usually result in the overall decrease in the fidelity without a significant 
change in the ripple. In any case, it is assuring that the overall performance does not bare a drastic dependence on the chosen operating 
parameters. Especially, a 5\% mismatch from optimal parameters leads to tolerable implications.

\subsection{$N=4$ cat-state generation}
Now, we would like to investigate the generation of the so-called, $N=4$ cat state \cite{mirrahimi14,albert16,roy16} 
by superposing equator- and polar-bound cat states. For this purpose, we can start with either of these cat states (production 
of which requires one pulse), and through a second pulse rotate it by $\pi/2$ back on to 
the $x$-axis, reproducing the $N=2$ cat state $\left[ |X\rangle +|-X\rangle\right]$ with a high fidelity. Then, under $\hat{H}_\eta$,
the time evolution of these antipodal CSSs will go through the squeezing stage, much like their isolated cases, apart from some
interference terms. Finally, applying a third rotation pulse (optimized in time and angle) around the $x$-axis will split and place 
one of them to the poles and the other to $\pm y$-axes, generating a $N=4$ state with a cross-legged cat construction of the target template 
$\left[ (|Z\rangle +|-Z\rangle)-(|Y\rangle +i|-Y\rangle) \right]$.

\begin{figure}
 \includegraphics[width=0.85\columnwidth]{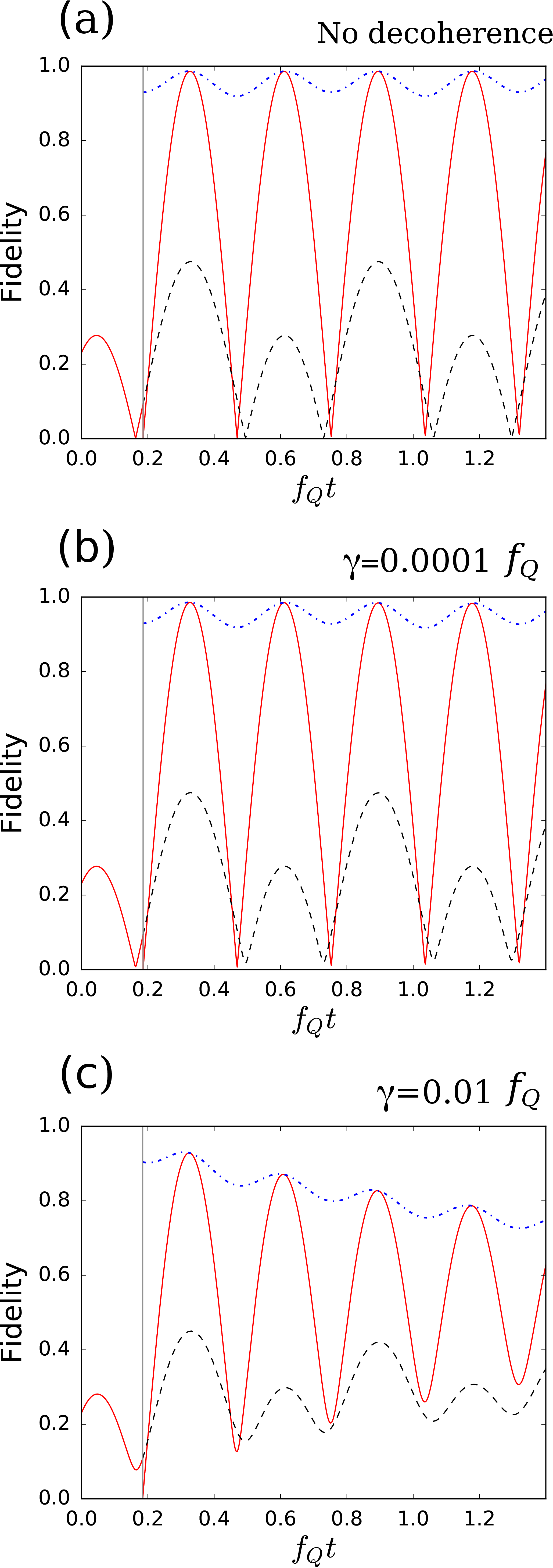}
\caption{Fidelity evolution without (dashed/black) and with (solid/red) rotation pulse, the latter resulting in a
 $N=4$ cat state for $\eta=1$ case of $I=5/2$. Also shown is the fidelity with respect to a one-leg-rotating target 
(dash-dot/blue) described by Eq.~(\ref{rot-target}). Vertical gray line marks the instant of the rotation pulse.
(a) Without any decoherence included, (b) and (c) with a phase damping rates of $\gamma=10^{-4}f_Q$ and $10^{-2}f_Q$, 
respectively.
}
\label{fig5}
\end{figure}

Figure~\ref{fig5} illustrates the fidelity with respect to this target state for the $\eta=1$ case of $I=5/2$. Here, almost full swing
oscillations are observed at an angular frequency of $\omega_2=2\pi\left(4\sqrt{7}f_Q/3\right)$. The time-evolving 
$N=4$ state can indeed be approximately represented by a rotating equator-bound 
cat state  (dashed lines in Fig.~\ref{fig5}) with respect to a polar-bound one in the form of
\begin{equation}
\label{rot-target}
\left[ (|Z\rangle +|-Z\rangle)+e^{i\omega_2 t}(|Y\rangle +i|-Y\rangle) \right]\, .
\end{equation}

\begin{figure}
 \includegraphics[width=\columnwidth]{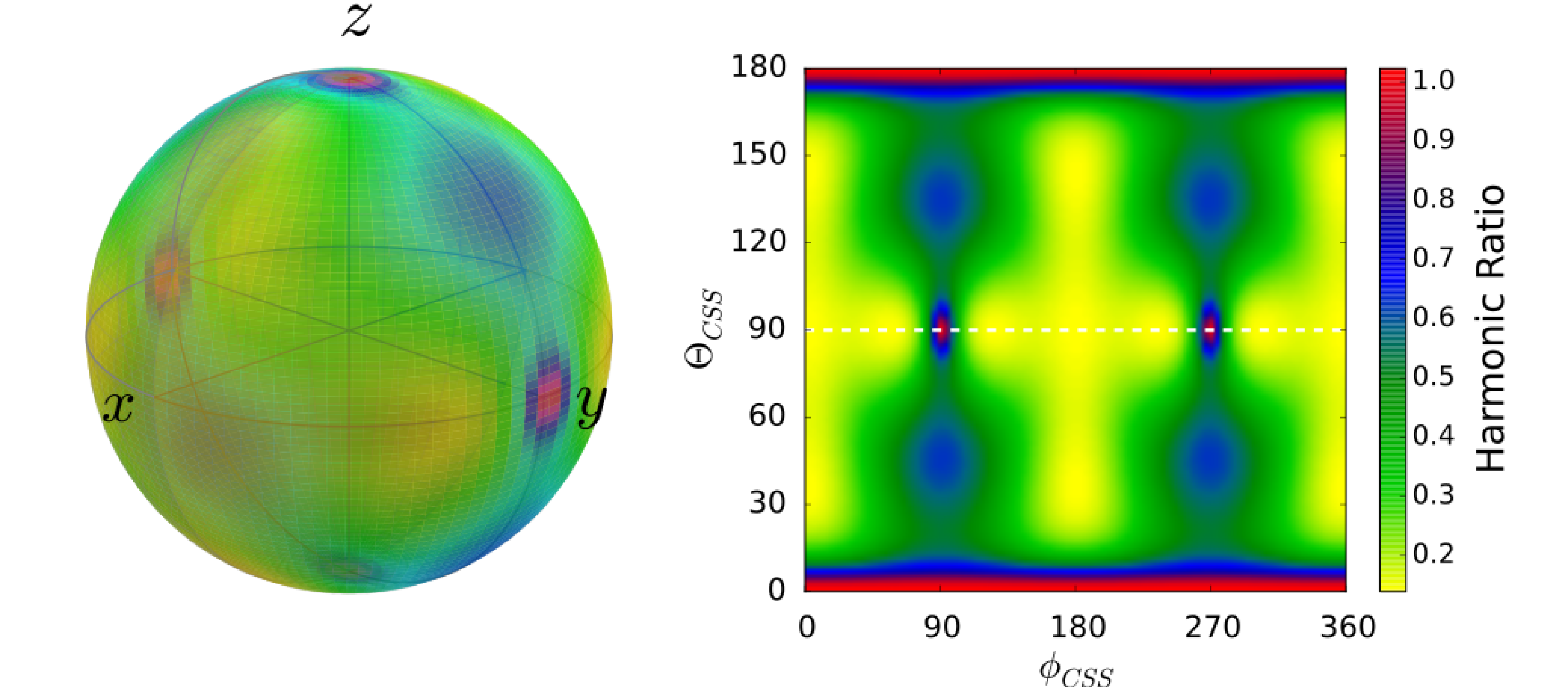}
\caption{The ratio of the second harmonic content with respect to fundamental in the evolution of fidelity 
as a function of the initial CSS location over the Bloch sphere for $I=5/2$.
}
\label{fig6}
\end{figure}

The Hamiltonian in Eq.~(\ref{H-QI}) for $\eta=1$ which corresponds to two-axis 
countertwisting has recently been shown to be amenable for a closed-form solution up to $I=21/2$ \cite{bhattacharya15}. 
Hence, our preference for the $I=5/2$ system is due to its strictly 
periodic (as opposed to quasi-periodic) time evolution \cite{korkmaz16}, stemming from two zero eigenfrequencies, 
and the other two at $\pm\omega_1=2\pi\left(2\sqrt{7}f_Q/3\right)$ \cite{bhattacharya15}. The intriguing point here is that the 
fidelity oscillation of the $N=4$ state occurs at its second harmonic, $\omega_2=2\omega_1$. 
In the Appendix we give the details on obtaining the explicit form of the 
time evolution operator, where it is shown that the doubly degenerate spectrum is responsible for the 
strong second harmonic content. This is observed, for instance in the time evolution of fidelity by the solid/red 
line in Fig.~\ref{fig5}, compared to the no-rotation-pulse case that also displays the fundamental frequency $\omega_1$. 
The ratio of second harmonic to fundamental under $\hat{H}_{\eta=1}$ depends on the location of the initial CSS over the the Bloch sphere, 
which is depicted in Fig.~\ref{fig6}.
In fact, this ratio becomes unity (actually meaning a frequency doubling) for CSS launched from $\left|\pm Y\right\rangle$ or 
$\left|\pm Z\right\rangle$ both coinciding with the countertwisting axes of $\hat{H}_1$. Therefore, we have $\omega_2=2\omega_1$ 
as the rotation frequency within the $N=4$ components of the target state. From a practical point of view, this internal rotation
offers an additional phase that can be benefited as an extra degree of freedom \cite{albert16}.

\subsection{Decoherence in cat states}
So far, our treatment was rather ideal other than considering the parameter sensitivity of our cat-state generation protocol.
Now, we would like to address the question of how good this recipe is in terms of practical realizability, 
predominantly when it is treated as an open system subject to environmental decoherence. As a matter of fact, from an 
experimental perspective it is well known that maintaining quantum coherence becomes exceedingly challenging as the distance 
between the superposed components of the cat state is increased \cite{brune96}. One advantage of nuclear-spin systems 
is their immunity to dissipative channels of spontaneous emission at radio frequencies \cite{gerry-knight} 
and the particle loss in contrast to cat states produced by Bose-Einstein condensates 
\cite{pezze16,byrnes12}, or cavity or circuit quantum electrodynamics \cite{haroche06,wallraff04}. 

In the presence of decoherence, the objective is to assure that the coupling 
of the nuclear spin to the intended degree of freedom is stronger than that of the dominant environmental process 
\cite{muller14}. In our context these are the quadrupolar frequency $f_Q$ versus the damping rate $\gamma$. 
The prevalent channel for the latter, in neutral solid-state spin systems (i.e., free from hyperfine coupling 
to the confined electronic spin) is the phase damping \cite{wust16}. For quantum dot structures of quadrupolar nuclei 
(e.g., $^{69,71}$Ga, $^{75}$As, $^{115}$In) the dephasing times ($T_2=1/\gamma$) lie in the 1--5~ms range 
\cite{chekhovich15,waeber16,wust16}. 
For the same systems the quadrupolar frequency dictated by strain is typically in the range $f_Q\sim$ 
2--8~MHz \cite{bulutay12,kuznetsova14,munsch14}. In the case of NV defect centers, the quadrupolar
$^{14}$N nuclear-spin dephasing times are at least, 1~ms \cite{ajoy12}, and the extracted $f_Q$ value is about 
10~MHz \cite{he93,ajoy12}. Thus, these two markedly distinct systems share highly similar values for 
$f_Q/\gamma=f_Q T_2\sim 10^{3}-10^{4}$, suggesting that strong quadrupolar coupling is attainable 
for such nuclear spins. We should note that there exist solid-state systems with even superior immunity to 
decoherence such as the single-crystal KClO$_3$ that has $f_Q=28.1$~MHz and $T_2=4.6$~ms with the 
product $f_Q T_2>10^{5}$ \cite{teles15}. As a caveat, if a nearby unpaired electron spin is present during the 
stabilization stage, it can degrade nuclear spin coherence \cite{neumann10}.

\begin{figure*}
\includegraphics[width=1.7\columnwidth]{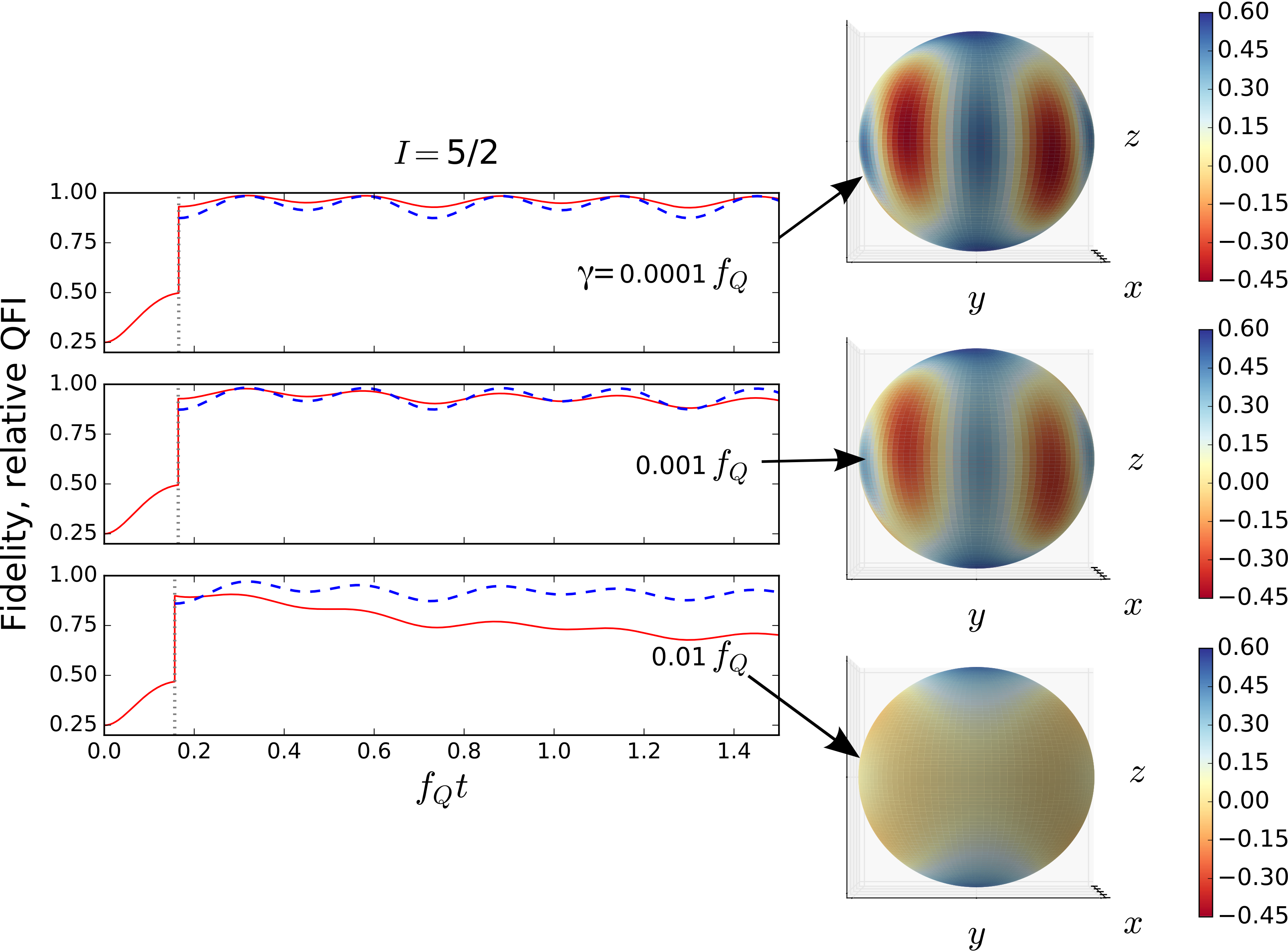}
\caption{For $I=5/2$, $\eta=1$, and polar-bound $N=2$ cat state the effect of decoherence. Left panel: Fidelity (solid/red) and 
normalized rQFI given by Eq.~(\ref{rQFI}) (dashed/blue). Right panel: Spin Wigner quasi-probability distributions at $10 t_R$.
Three different dephasing rates are considered, $\gamma/f_Q=10^{-4},\, 10^{-3},\, 10^{-2}$ from top to bottom, respectively.
}
\label{fig7}
\end{figure*}  

In the following, we consider dephasing rates $\gamma/f_Q$ ranging from $10^{-4}$ to $10^{-2}$, which allows for 
harsher decoherence to observe its adverse consequences. In Fig.~\ref{fig7} we display how fidelity and rQFI
of polar-bound $N=2$ cat states are affected from decoherence for $I=5/2$ and $\eta=1$. The Wigner distributions on the 
right panel refer to $t=10t_R$, i.e., at 10 times the rotation pulse instant, $t_R$. 
As would be anticipated from the strong quadrupolar regime, the case for $\gamma=10^{-4}f_Q$ is virtually indistinguishable
within this time frame from the decoherence-free ones in Fig.~\ref{fig2}(d) and (g). 
For $\gamma=10^{-3}f_Q $ the deviation becomes noticeable, and this gets drastic for $\gamma=10^{-2}f_Q$, especially 
on the fidelity, whereas the normalized rQFI measure (see, Eq.~(\ref{rQFI})) is much less affected as it is insensitive to 
phase coherence, and rather probes the separation of the CSSs. The Wigner distributions are also instrumental 
in tracking these differences through the attained 
negative values \cite{byrnes16}, which are in general taken as a measure of the quantumness of the states \cite{agarwal97}. 
It is clearly seen that decoherence gradually removes these negative interference fringes of the superposition. 
We note that the equator-bound $N=2$ cat states (not shown) are somewhat less susceptible than the polar-bound ones.

In the same vein, we return to Fig.~\ref{fig5} to discuss how decoherence affects $N=4$ cat states, which reassures us that
a rate of $\gamma=10^{-4}f_Q$ is not influential whereas $10^{-2}f_Q$ becomes very destructive 
on the fidelity by washing away the contrast between originally orthogonal states.
Here, the point to note is that $N=4$ cat states do not particularly suffer more from decoherence than $N=2$ variants.
Based on these insights we can conclude that the proposed scheme is decoherence tolerant for rates around 
$\gamma/f_Q\sim 10^{-4}$, that is, in the upper end of currently available range without any additional countermeasures 
such as the dynamical decoupling of the environmental spins \cite{suter16}.

\begin{figure}
 \includegraphics[width=\columnwidth]{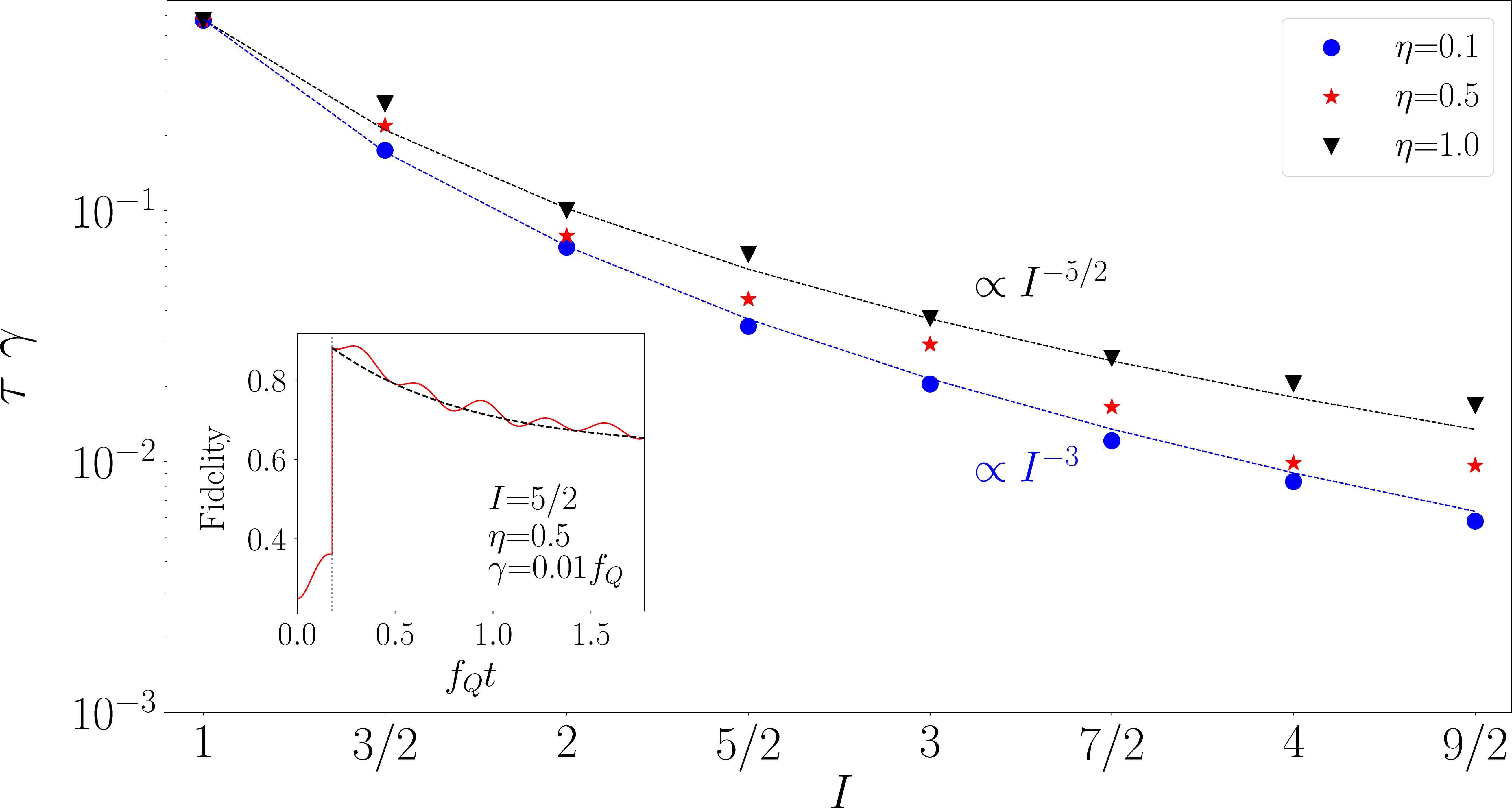}
\caption{The variation of fidelity decay time constant, $\tau$ (in units of reciprocal dephasing rate, $1/\gamma$) with spin-$I$.
Polar-bound $N=2$ cat states are considered. Dashed lines mark the $I^{-5/2}$ and $I^{-3}$ scalings. 
Inset shows the damping of fidelity after the rotation pulse for
the case of $I=5/2$, $\eta=0.5$, and $\gamma=10^{-2}f_Q$; dashed line is the exponential fit to extract the decay time constant.
}
\label{fig8}
\end{figure}

According to the qudit dephasing model employed in this work, the number of channels increases in proportion to spin 
angular momentum $I$, as can be seen from Eqs.~(\ref{lindblad-eq}) and (\ref{lindblad-op}). Furthermore, we are
dealing with quite unique macroscopic quantum spin states that are not necessarily governed by the same dephasing rate 
$\gamma$ which applies to Dicke states \cite{haroche06}. Therefore, we need to identify how the 
baseline dephasing rate, $\gamma$ compares with the decay rate of cat state fidelities as a function of $I$. The inset in
Fig.~\ref{fig8} displays a typical damping of fidelity under decoherence (here, for $\gamma=10^{-2}f_Q$) which has an 
oscillatory pattern for the specific $\eta=0.5$ and $I=5/2$ values considered. Its time constant $\tau$ can be extracted 
by fitting the fidelity to a form $F(t)=F_0\exp\left(-t/\tau\right)+F_{\rm sat}$. Figure~\ref{fig8} illustrates the 
scaling of the fidelity decay time constant as a function of $I$ for the $N=2$ cat states. It reveals that at the lower end of $I$ 
the fidelity decay rate ($1/\tau$) approaches toward $\gamma$. As $I$ increases, its scaling lies roughly in between 
$I^3$ and $I^{5/2}$, for $\eta =0.1$ and $\eta =1$, respectively. The overall behavior is essentially 
preserved for different baseline dephasing rates in the range $\gamma=10^{-2}-10^{-4}$. The enhanced susceptibility to decoherence
of the large-$I$ spin cat states is already known, like in the context of spinor Bose-Einstein condensates \cite{byrnes15}.
A trivial advantage of nuclear spins is that the $I$ values are inherently capped before this scaling becomes prohibitive.
Therefore, we believe that there exits opportunities for spin cat states within this range of low angular momenta.

\subsection{Discussions on practical aspects}
Finally, we would like to address key practical issues in the realization of this theoretical proposal 
in regard to initialization, manipulation, and readout. As the exemplary platform we have particularly in mind 
the NV centers \cite{muller14}.
We begin with the initialization of the nuclear spin state. Our aforementioned cat-state generation schemes start 
from the $|+X\rangle$ CSS which is accessible from, say, the $\left|I, I\right\rangle$ Dicke state by a $\pi/2$ rotation. 
In negatively charged defect centers, a nuclear spin state like $\left|I, I\right\rangle$ can be initialized 
with high fidelity by transferring the polarization from the ancillary electron spin \cite{waldherr14}. 
The latter is easily obtained by means of optically induced electronic spin orientation \cite{manson06}. 
Most commonly the transfer of polarization from the electronic to nuclear spin is mediated by the hyperfine 
interaction \cite{smeltzer09}. But, alternatively using a forbidden transition with a two-photon process was also 
proposed \cite{ajoy12}.

After initialization and a brief interval of QI evolution, the rotation around one of the principal EFG axes 
is the next manipulation to be imposed on the nuclear spin. Traditionally rotation is accomplished by Rabi 
flopping, through a resonant rf pulse of amplitude and duration to satisfy the pulse area for the 
desired rotation angle \cite{gerry-knight}. The nuclear magnetic resonance community has developed measures to 
attain the intended rotation in the presence of systematic frequency or pulse duration errors \cite{cummins03}.
A different choice is to rotate the nuclear spin via off-resonant Rabi oscillation of the electron 
spin and the hyperfine interaction \cite{morton06}. This has been experimentally tested on NV centers to yield 
about 2 orders of magnitude shorter rotation duration compared to the traditional approach with moderate pulse 
amplitudes \cite{sangtawesin14}. As a matter of fact, fast spin rotation is highly desirable for our framework where we 
assume it to be instantaneous. Indeed, a few hundred nanosecond nuclear spin rotation timescale \cite{sangtawesin14} 
would meet this requirement against microsecond-long durations of QI.

The final stage is the detection, where a projective measurement is performed conventionally on the observable $I_z$.
However, with prior rotations $I_x$ or $I_y$ can likewise be read out. As a matter of fact, using a sequence of 
rotation pulses, measuring the free-induction decay signals, and Fourier transforming them forms the recipe to extract
all entries of the density matrix, with real and imaginary parts \cite{chuang98}. This is known as quantum state tomography,
which has been in use for the nuclear spins for the past two decades; as a relevant example the density matrix reconstruction
of three entangled nuclear spins with a nondemolition readout in NV centers can be given \cite{waldherr14,cramer16}. 
Moreover, state tomography has been extended to the quadrupolar nuclei \cite{teles07}, and there are also efforts to implement 
it with small or zero static magnetic field, which in effect makes it the nuclear quadrupole resonance \cite{teles15}.
In general, once the full density matrix is at one's disposal by state tomography, the Wigner distributions as in this work, 
can be directly computed for target state comparison; a recent such demonstration combining experiment is given in 
Ref.~\cite{auccaise15}. Depending on the encoding strategy, rather than the projective $I_z$ 
measurement, a different detection, like the parity of the superposed cat states is another option \cite{rundle16}.

It needs to be mentioned that the involvement of neighboring spins, including the ancillary electron spin, albeit with
all its aforestated advantages, gives rise to measurement errors \cite{neumann10,robledo11,pla13,dreau13}. 
Specifically in the case of electron spin,
the component of the hyperfine tensor perpendicular to the NV axis causes stochastic quantum jumps in the nuclear spin. 
Therefore, one should assure the mean jump interval to be longer than the measurement time \cite{neumann10}.

An important matter that remains is the role of entanglement. The Schr\"{o}dinger cat states 
have long been advocated as the key enabler for quantum sensing to surpass the standard 
quantum limit \cite{pezze16}. The common framework, as in two-component Bose-Einstein condensates, hinges upon $N$-particle 
entanglement in constructing a spin-$N/2$ collective Hilbert space \cite{byrnes12}. On the other hand, for the paradigm 
considered in this work, the entanglement of nuclei plays no role, as it is based on the pure state of 
a {\em single} nuclear spin \cite{gedik15}. The fact that it is a quadrupolar nucleus avails a qudit structure, 
also granting its controllability via the quadratic Hamiltonian of Eq.~(\ref{H-QI}) that facilitates 
squeezing \cite{korkmaz16}, and cat-state generation as in this work. Achieving all of these without a need for entanglement
with other nuclei makes it less fragile under decoherence, as elucidated in the above analysis. Such an ability to generate 
and control cat states on a nuclear spin with $I\ge 1$ amounts to a small-scale quantum information processor, much like other 
prototypes that can be put to use in various ways \cite{steffen16}. For expanding our approach, the route of 
exploiting yet higher level of superpositions (i.e., $N>4$) of the cat states while remaining in the same spin angular 
momentum subspace is blocked because of the limited number of fixed points over the Bloch sphere of the Hamiltonian 
(in our case, QI). Hence, if the scalability is the primary objective one should bring in an additional 
layer of spin-spin interaction among nuclei to harness entanglement through an enlarged Hilbert space \cite{leuenberger03}.

\section{Conclusions}
In summary, we present a blueprint for generating stabilized nuclear spin cat states using biaxial QI together 
with one or three rotation pulses. A rudimentary optimization of the single-pulse approach attains fidelities 
around 0.95, while being largely insensitive to the variations in the parameters. After analyzing the polar- 
and equator-bound $N=2$ cat states separately, we considered their superposition with four CSSs,
where one of the two-component cat states rotates with respect to the EFG axes. Such states play a crucial 
role in cat codes to protect against bit flips \cite{ofek16}. To render our analysis more realistic, the effect 
of phase noise, which is the dominant decoherence mechanism, is thoroughly investigated, showing that these 
generated cat states can retain their fidelities on the favorable end of the currently accessible decoherence levels.

% Suggestions for future work
We believe that optically addressable color centers involving a quadrupolar nucleus, like an implanted indium defect
within a wide-bandgap-host \cite{hausmann11}, appears to be a suitable physical system for realizing nuclear-spin cat states. 
The other option of self-assembled quantum dots possesses an {\em ensemble} of more than ten thousands of quadrupolar 
nuclei that can be to some extent entangled though the confined electron spin \cite{yang09,stanek14}. The primary 
challenge here is the spatially inhomogeneous strain causing the tilting variations of EFG axes within a solid-state matrix 
\cite{bulutay12,bulutay14}. Therefore, a means for narrowing this distribution may be valuable to 
gain better control over this resource.
As other possible extensions, schemes for further increasing the maximum fidelity together 
with low ripple can be sought to meet the stringent practical demands \cite{knill05}. This is particularly relevant 
for the recently introduced cat codes utilizing microwave photons in superconducting circuits, which 
have proven to be practical for quantum error correction \cite{ofek16}. 
Therefore, their nuclear-spin cat-state implementation may be pursued both theoretically and experimentally.

\begin{acknowledgments}
We are grateful to Tom\'{a}\v{s} Opatrn\'{y} for fruitful discussions.
This work was supported by T\"UB\.ITAK, The Scientific and Technological Research Council 
of Turkey through the project No. 114F409.
The numerical calculations reported in this paper were partially performed at T\"UB\.ITAK ULAKB\.IM, 
High Performance and Grid Computing Center (TRUBA resources).
\end{acknowledgments}

\appendix
\section*{Appendix: Time evolution operator for spin-5/2}
In this section our aim is to obtain the closed form expression for the time evolution operator under the QI Hamiltonian 
of a spin-5/2 system, specifically for $\eta=1$. The Hamiltonian in Eq.~(\ref{H-QI}) reduces in this case to 
\begin{equation}
\label{H-TAC}
\hat{H}_{\eta=1} = \frac{\hbar \omega_Q}{3}\left( \hat{I}_z^2 - \hat{I}_y^2\right)\, ,
\end{equation}
where, $\omega_Q = 2\pi f_Q$, and the associated characteristic polynomial for $H_1$ is given by \cite{bhattacharya15},
\begin{equation}
p_{H_1}(\lambda) = \lambda^2 \left( \lambda^2 - 28\right)^2 \, .
\end{equation}
Thus, the six roots are composed of one at zero, and two equal in magnitude but opposite sign eigenfrequencies, each of them 
being doubly degenerate. That is, the distinct spectrum is composed of $\lambda_j=\{0,\, \omega_1,\, -\omega_1\}$, 
with $\omega_1=2\sqrt{7}\omega_Q/3$. 
Taking into account the degeneracies in the spectrum \cite{dezela14}, we can work out the time evolution operator under
$\hat{H}_1$ explicitly, starting from
\begin{equation}
e^{-i\hat{H}_1 t/\hbar} = \sum\limits_{j=1}^3 e^{-i\lambda_j t} \prod\limits_{\buildrel {k=1}\over {(k\ne j)} }^3 
\frac{i\hat{H}_1 t-\lambda_k \hat{1}}{\lambda_j - \lambda_k} \, ,
\end{equation}
where $\hat{1}$ is the identity operator. After inserting the eigenfrequencies it leads to the following closed form expression, 
\begin{equation}
\label{TAC-evolve}
e^{-i\hat{H}_1 t/\hbar} =  \frac{\cos\left(\omega_1 t\right) - 1}{\omega_1^2} \hat{H}_1^2 - \frac{i\sin\left(\omega_1 t\right) }{\omega_1} \hat{H}_1 + \hat{1}\, .
\end{equation}
As mentioned in the main text, the double degeneracy in the spectrum gives rise to second-harmonic generation with respect to the 
fundamental eigenfrequency of $\omega_1$.


\begin{thebibliography}{87}%
\makeatletter
\providecommand \@ifxundefined [1]{%
 \@ifx{#1\undefined}
}%
\providecommand \@ifnum [1]{%
 \ifnum #1\expandafter \@firstoftwo
 \else \expandafter \@secondoftwo
 \fi
}%
\providecommand \@ifx [1]{%
 \ifx #1\expandafter \@firstoftwo
 \else \expandafter \@secondoftwo
 \fi
}%
\providecommand \natexlab [1]{#1}%
\providecommand \enquote  [1]{``#1''}%
\providecommand \bibnamefont  [1]{#1}%
\providecommand \bibfnamefont [1]{#1}%
\providecommand \citenamefont [1]{#1}%
\providecommand \href@noop [0]{\@secondoftwo}%
\providecommand \href [0]{\begingroup \@sanitize@url \@href}%
\providecommand \@href[1]{\@@startlink{#1}\@@href}%
\providecommand \@@href[1]{\endgroup#1\@@endlink}%
\providecommand \@sanitize@url [0]{\catcode `\\12\catcode `\$12\catcode
  `\&12\catcode `\#12\catcode `\^12\catcode `\_12\catcode `\%12\relax}%
\providecommand \@@startlink[1]{}%
\providecommand \@@endlink[0]{}%
\providecommand \url  [0]{\begingroup\@sanitize@url \@url }%
\providecommand \@url [1]{\endgroup\@href {#1}{\urlprefix }}%
\providecommand \urlprefix  [0]{URL }%
\providecommand \Eprint [0]{\href }%
\providecommand \doibase [0]{http://dx.doi.org/}%
\providecommand \selectlanguage [0]{\@gobble}%
\providecommand \bibinfo  [0]{\@secondoftwo}%
\providecommand \bibfield  [0]{\@secondoftwo}%
\providecommand \translation [1]{[#1]}%
\providecommand \BibitemOpen [0]{}%
\providecommand \bibitemStop [0]{}%
\providecommand \bibitemNoStop [0]{.\EOS\space}%
\providecommand \EOS [0]{\spacefactor3000\relax}%
\providecommand \BibitemShut  [1]{\csname bibitem#1\endcsname}%
\let\auto@bib@innerbib\@empty
%</preamble>
\bibitem [{\citenamefont {Dowling}\ and\ \citenamefont
  {Milburn}(2003)}]{dowling03}%
  \BibitemOpen
  \bibfield  {author} {\bibinfo {author} {\bibfnamefont {J.~P.}\ \bibnamefont
  {Dowling}}\ and\ \bibinfo {author} {\bibfnamefont {G.~J.}\ \bibnamefont
  {Milburn}},\ }\bibfield  {title} {\enquote {\bibinfo {title} {Quantum
  technology: the second quantum revolution},}\ }\href@noop {} {\bibfield
  {journal} {\bibinfo  {journal} {Philos. Trans. R. Soc. London, Ser. A}\
  }\textbf {\bibinfo {volume} {361}},\ \bibinfo {pages} {1655--1674} (\bibinfo
  {year} {2003})}\BibitemShut {NoStop}%
\bibitem [{\citenamefont {Cory}\ \emph {et~al.}(1997)\citenamefont {Cory},
  \citenamefont {Fahmy},\ and\ \citenamefont {Havel}}]{cory97}%
  \BibitemOpen
  \bibfield  {author} {\bibinfo {author} {\bibfnamefont {D.~G.}\ \bibnamefont
  {Cory}}, \bibinfo {author} {\bibfnamefont {A.~F.}\ \bibnamefont {Fahmy}}, \
  and\ \bibinfo {author} {\bibfnamefont {T.~F.}\ \bibnamefont {Havel}},\
  }\bibfield  {title} {\enquote {\bibinfo {title} {Ensemble quantum computing
  by nmr spectroscopy},}\ }\href@noop {} {\bibfield  {journal} {\bibinfo
  {journal} {Proc. Natl. Acad. Sci. U.S.A}\ }\textbf {\bibinfo {volume} {94}},\
  \bibinfo {pages} {1634--1639} (\bibinfo {year} {1997})}\BibitemShut {NoStop}%
\bibitem [{\citenamefont {Kane}(1998)}]{kane98}%
  \BibitemOpen
  \bibfield  {author} {\bibinfo {author} {\bibfnamefont {B.~E.}\ \bibnamefont
  {Kane}},\ }\bibfield  {title} {\enquote {\bibinfo {title} {A silicon-based
  nuclear spin quantum computer},}\ }\href@noop {} {\bibfield  {journal}
  {\bibinfo  {journal} {Nature (London)}\ }\textbf {\bibinfo {volume} {393}},\
  \bibinfo {pages} {133--137} (\bibinfo {year} {1998})}\BibitemShut {NoStop}%
\bibitem [{\citenamefont {Levitt}(2007)}]{levitt07}%
  \BibitemOpen
  \bibfield  {author} {\bibinfo {author} {\bibfnamefont {M.~H.}\ \bibnamefont
  {Levitt}},\ }\href@noop {} {\emph {\bibinfo {title} {Spin dynamics: basics of
  nuclear magnetic resonance}}}\ (\bibinfo  {publisher} {John Wiley \& Sons},\
  \bibinfo {year} {2007})\BibitemShut {NoStop}%
\bibitem [{\citenamefont {Warren}(1997)}]{warren97}%
  \BibitemOpen
  \bibfield  {author} {\bibinfo {author} {\bibfnamefont {W.~S.}\ \bibnamefont
  {Warren}},\ }\bibfield  {title} {\enquote {\bibinfo {title} {The usefulness
  of nmr quantum computing},}\ }\href@noop {} {\bibfield  {journal} {\bibinfo
  {journal} {Science}\ }\textbf {\bibinfo {volume} {277}},\ \bibinfo {pages}
  {1688--1690} (\bibinfo {year} {1997})}\BibitemShut {NoStop}%
\bibitem [{\citenamefont {K{\"o}hler}\ \emph {et~al.}(1993)\citenamefont
  {K{\"o}hler}, \citenamefont {Disselhorst}, \citenamefont {Donckers},
  \citenamefont {Groenen}, \citenamefont {Schmidt},\ and\ \citenamefont
  {Moerner}}]{kohler93}%
  \BibitemOpen
  \bibfield  {author} {\bibinfo {author} {\bibfnamefont {J.}~\bibnamefont
  {K{\"o}hler}}, \bibinfo {author} {\bibfnamefont {J.~A. J.~M.}\ \bibnamefont
  {Disselhorst}}, \bibinfo {author} {\bibfnamefont {M.~C. J.~M.}\ \bibnamefont
  {Donckers}}, \bibinfo {author} {\bibfnamefont {E.~J.~J.}\ \bibnamefont
  {Groenen}}, \bibinfo {author} {\bibfnamefont {J.}~\bibnamefont {Schmidt}}, \
  and\ \bibinfo {author} {\bibfnamefont {W.~E.}\ \bibnamefont {Moerner}},\
  }\bibfield  {title} {\enquote {\bibinfo {title} {Magnetic resonance of a
  single molecular spin},}\ }\href@noop {} {\bibfield  {journal} {\bibinfo
  {journal} {Nature (London)}\ }\textbf {\bibinfo {volume} {363}},\ \bibinfo
  {pages} {242--244} (\bibinfo {year} {1993})}\BibitemShut {NoStop}%
\bibitem [{\citenamefont {Wrachtrup}\ \emph {et~al.}(1993)\citenamefont
  {Wrachtrup}, \citenamefont {Von~Borczyskowski}, \citenamefont {Bernard},
  \citenamefont {Orritt},\ and\ \citenamefont {Brown}}]{wrachtrup93}%
  \BibitemOpen
  \bibfield  {author} {\bibinfo {author} {\bibfnamefont {J.}~\bibnamefont
  {Wrachtrup}}, \bibinfo {author} {\bibfnamefont {C.}~\bibnamefont
  {Von~Borczyskowski}}, \bibinfo {author} {\bibfnamefont {J.}~\bibnamefont
  {Bernard}}, \bibinfo {author} {\bibfnamefont {M.}~\bibnamefont {Orritt}}, \
  and\ \bibinfo {author} {\bibfnamefont {R.}~\bibnamefont {Brown}},\ }\bibfield
   {title} {\enquote {\bibinfo {title} {Optical detection of magnetic resonance
  in a single molecule},}\ }\href@noop {} {\bibfield  {journal} {\bibinfo
  {journal} {Nature (London)}\ }\textbf {\bibinfo {volume} {363}},\ \bibinfo
  {pages} {244--245} (\bibinfo {year} {1993})}\BibitemShut {NoStop}%
\bibitem [{\citenamefont {Wrachtrup}\ \emph {et~al.}(1997)\citenamefont
  {Wrachtrup}, \citenamefont {Gruber}, \citenamefont {Fleury},\ and\
  \citenamefont {Von~Borczyskowski}}]{wrachtrup97}%
  \BibitemOpen
  \bibfield  {author} {\bibinfo {author} {\bibfnamefont {J.}~\bibnamefont
  {Wrachtrup}}, \bibinfo {author} {\bibfnamefont {A.}~\bibnamefont {Gruber}},
  \bibinfo {author} {\bibfnamefont {L.}~\bibnamefont {Fleury}}, \ and\ \bibinfo
  {author} {\bibfnamefont {C.}~\bibnamefont {Von~Borczyskowski}},\ }\bibfield
  {title} {\enquote {\bibinfo {title} {Magnetic resonance on single nuclei},}\
  }\href@noop {} {\bibfield  {journal} {\bibinfo  {journal} {Chem. Phys.
  Lett.}\ }\textbf {\bibinfo {volume} {267}},\ \bibinfo {pages} {179--185}
  (\bibinfo {year} {1997})}\BibitemShut {NoStop}%
\bibitem [{\citenamefont {Jelezko}\ \emph {et~al.}(2004)\citenamefont
  {Jelezko}, \citenamefont {Gaebel}, \citenamefont {Popa}, \citenamefont
  {Domhan}, \citenamefont {Gruber},\ and\ \citenamefont
  {Wrachtrup}}]{jelezko04b}%
  \BibitemOpen
  \bibfield  {author} {\bibinfo {author} {\bibfnamefont {F.}~\bibnamefont
  {Jelezko}}, \bibinfo {author} {\bibfnamefont {T.}~\bibnamefont {Gaebel}},
  \bibinfo {author} {\bibfnamefont {I.}~\bibnamefont {Popa}}, \bibinfo {author}
  {\bibfnamefont {M.}~\bibnamefont {Domhan}}, \bibinfo {author} {\bibfnamefont
  {A.}~\bibnamefont {Gruber}}, \ and\ \bibinfo {author} {\bibfnamefont
  {J.}~\bibnamefont {Wrachtrup}},\ }\bibfield  {title} {\enquote {\bibinfo
  {title} {Observation of coherent oscillation of a single nuclear spin and
  realization of a two-qubit conditional quantum gate},}\ }\href {\doibase
  10.1103/PhysRevLett.93.130501} {\bibfield  {journal} {\bibinfo  {journal}
  {Phys. Rev. Lett.}\ }\textbf {\bibinfo {volume} {93}},\ \bibinfo {pages}
  {130501} (\bibinfo {year} {2004})}\BibitemShut {NoStop}%
\bibitem [{\citenamefont {Dutt}\ \emph {et~al.}(2007)\citenamefont {Dutt},
  \citenamefont {Childress}, \citenamefont {Jiang}, \citenamefont {Togan},
  \citenamefont {Maze}, \citenamefont {Jelezko}, \citenamefont {Zibrov},
  \citenamefont {Hemmer},\ and\ \citenamefont {Lukin}}]{dutt07}%
  \BibitemOpen
  \bibfield  {author} {\bibinfo {author} {\bibfnamefont {M.~V.~G.}\
  \bibnamefont {Dutt}}, \bibinfo {author} {\bibfnamefont {L.}~\bibnamefont
  {Childress}}, \bibinfo {author} {\bibfnamefont {L.}~\bibnamefont {Jiang}},
  \bibinfo {author} {\bibfnamefont {E.}~\bibnamefont {Togan}}, \bibinfo
  {author} {\bibfnamefont {J.}~\bibnamefont {Maze}}, \bibinfo {author}
  {\bibfnamefont {F.}~\bibnamefont {Jelezko}}, \bibinfo {author} {\bibfnamefont
  {A.~S.}\ \bibnamefont {Zibrov}}, \bibinfo {author} {\bibfnamefont {P.~R.}\
  \bibnamefont {Hemmer}}, \ and\ \bibinfo {author} {\bibfnamefont {M.~D.}\
  \bibnamefont {Lukin}},\ }\bibfield  {title} {\enquote {\bibinfo {title}
  {Quantum register based on individual electronic and nuclear spin qubits in
  diamond},}\ }\href@noop {} {\bibfield  {journal} {\bibinfo  {journal}
  {Science}\ }\textbf {\bibinfo {volume} {316}},\ \bibinfo {pages} {1312--1316}
  (\bibinfo {year} {2007})}\BibitemShut {NoStop}%
\bibitem [{\citenamefont {Neumann}\ \emph {et~al.}(2010)\citenamefont
  {Neumann}, \citenamefont {Beck}, \citenamefont {Steiner}, \citenamefont
  {Rempp}, \citenamefont {Fedder}, \citenamefont {Hemmer}, \citenamefont
  {Wrachtrup},\ and\ \citenamefont {Jelezko}}]{neumann10}%
  \BibitemOpen
  \bibfield  {author} {\bibinfo {author} {\bibfnamefont {P.}~\bibnamefont
  {Neumann}}, \bibinfo {author} {\bibfnamefont {J.}~\bibnamefont {Beck}},
  \bibinfo {author} {\bibfnamefont {M.}~\bibnamefont {Steiner}}, \bibinfo
  {author} {\bibfnamefont {F.}~\bibnamefont {Rempp}}, \bibinfo {author}
  {\bibfnamefont {H.}~\bibnamefont {Fedder}}, \bibinfo {author} {\bibfnamefont
  {P.~R.}\ \bibnamefont {Hemmer}}, \bibinfo {author} {\bibfnamefont
  {J.}~\bibnamefont {Wrachtrup}}, \ and\ \bibinfo {author} {\bibfnamefont
  {F.}~\bibnamefont {Jelezko}},\ }\bibfield  {title} {\enquote {\bibinfo
  {title} {Single-shot readout of a single nuclear spin},}\ }\href@noop {}
  {\bibfield  {journal} {\bibinfo  {journal} {Science}\ }\textbf {\bibinfo
  {volume} {329}},\ \bibinfo {pages} {542--544} (\bibinfo {year}
  {2010})}\BibitemShut {NoStop}%
\bibitem [{\citenamefont {Robledo}\ \emph {et~al.}(2011)\citenamefont
  {Robledo}, \citenamefont {Childress}, \citenamefont {Bernien}, \citenamefont
  {Hensen}, \citenamefont {Alkemade},\ and\ \citenamefont
  {Hanson}}]{robledo11}%
  \BibitemOpen
  \bibfield  {author} {\bibinfo {author} {\bibfnamefont {L.}~\bibnamefont
  {Robledo}}, \bibinfo {author} {\bibfnamefont {L.}~\bibnamefont {Childress}},
  \bibinfo {author} {\bibfnamefont {H.}~\bibnamefont {Bernien}}, \bibinfo
  {author} {\bibfnamefont {B.}~\bibnamefont {Hensen}}, \bibinfo {author}
  {\bibfnamefont {P.~F.~A.}\ \bibnamefont {Alkemade}}, \ and\ \bibinfo {author}
  {\bibfnamefont {R.}~\bibnamefont {Hanson}},\ }\bibfield  {title} {\enquote
  {\bibinfo {title} {High-fidelity projective read-out of a solid-state spin
  quantum register},}\ }\href@noop {} {\bibfield  {journal} {\bibinfo
  {journal} {Nature (London)}\ }\textbf {\bibinfo {volume} {477}},\ \bibinfo
  {pages} {574--578} (\bibinfo {year} {2011})}\BibitemShut {NoStop}%
\bibitem [{\citenamefont {Dr\'eau}\ \emph {et~al.}(2013)\citenamefont
  {Dr\'eau}, \citenamefont {Spinicelli}, \citenamefont {Maze}, \citenamefont
  {Roch},\ and\ \citenamefont {Jacques}}]{dreau13}%
  \BibitemOpen
  \bibfield  {author} {\bibinfo {author} {\bibfnamefont {A.}~\bibnamefont
  {Dr\'eau}}, \bibinfo {author} {\bibfnamefont {P.}~\bibnamefont {Spinicelli}},
  \bibinfo {author} {\bibfnamefont {J.~R.}\ \bibnamefont {Maze}}, \bibinfo
  {author} {\bibfnamefont {J.-F.}\ \bibnamefont {Roch}}, \ and\ \bibinfo
  {author} {\bibfnamefont {V.}~\bibnamefont {Jacques}},\ }\bibfield  {title}
  {\enquote {\bibinfo {title} {Single-shot readout of multiple nuclear spin
  qubits in diamond under ambient conditions},}\ }\href {\doibase
  10.1103/PhysRevLett.110.060502} {\bibfield  {journal} {\bibinfo  {journal}
  {Phys. Rev. Lett.}\ }\textbf {\bibinfo {volume} {110}},\ \bibinfo {pages}
  {060502} (\bibinfo {year} {2013})}\BibitemShut {NoStop}%
\bibitem [{\citenamefont {Vincent}\ \emph {et~al.}(2012)\citenamefont
  {Vincent}, \citenamefont {Klyatskaya}, \citenamefont {Ruben}, \citenamefont
  {Wernsdorfer},\ and\ \citenamefont {Balestro}}]{vincent12}%
  \BibitemOpen
  \bibfield  {author} {\bibinfo {author} {\bibfnamefont {R.}~\bibnamefont
  {Vincent}}, \bibinfo {author} {\bibfnamefont {S.}~\bibnamefont {Klyatskaya}},
  \bibinfo {author} {\bibfnamefont {M.}~\bibnamefont {Ruben}}, \bibinfo
  {author} {\bibfnamefont {W.}~\bibnamefont {Wernsdorfer}}, \ and\ \bibinfo
  {author} {\bibfnamefont {F.}~\bibnamefont {Balestro}},\ }\bibfield  {title}
  {\enquote {\bibinfo {title} {Electronic read-out of a single nuclear spin
  using a molecular spin transistor},}\ }\href@noop {} {\bibfield  {journal}
  {\bibinfo  {journal} {Nature (London)}\ }\textbf {\bibinfo {volume} {488}},\
  \bibinfo {pages} {357--360} (\bibinfo {year} {2012})}\BibitemShut {NoStop}%
\bibitem [{\citenamefont {Pla}\ \emph {et~al.}(2013)\citenamefont {Pla},
  \citenamefont {Tan}, \citenamefont {Dehollain}, \citenamefont {Lim},
  \citenamefont {Morton}, \citenamefont {Zwanenburg}, \citenamefont {Jamieson},
  \citenamefont {Dzurak},\ and\ \citenamefont {Morello}}]{pla13}%
  \BibitemOpen
  \bibfield  {author} {\bibinfo {author} {\bibfnamefont {J.~J.}\ \bibnamefont
  {Pla}}, \bibinfo {author} {\bibfnamefont {K.~Y.}\ \bibnamefont {Tan}},
  \bibinfo {author} {\bibfnamefont {J.~P.}\ \bibnamefont {Dehollain}}, \bibinfo
  {author} {\bibfnamefont {W.~H.}\ \bibnamefont {Lim}}, \bibinfo {author}
  {\bibfnamefont {J.~J.~L.}\ \bibnamefont {Morton}}, \bibinfo {author}
  {\bibfnamefont {F.~A.}\ \bibnamefont {Zwanenburg}}, \bibinfo {author}
  {\bibfnamefont {D.~N.}\ \bibnamefont {Jamieson}}, \bibinfo {author}
  {\bibfnamefont {A.~S.}\ \bibnamefont {Dzurak}}, \ and\ \bibinfo {author}
  {\bibfnamefont {A.}~\bibnamefont {Morello}},\ }\bibfield  {title} {\enquote
  {\bibinfo {title} {High-fidelity readout and control of a nuclear spin qubit
  in silicon},}\ }\href@noop {} {\bibfield  {journal} {\bibinfo  {journal}
  {Nature (London)}\ }\textbf {\bibinfo {volume} {496}},\ \bibinfo {pages}
  {334--338} (\bibinfo {year} {2013})}\BibitemShut {NoStop}%
\bibitem [{\citenamefont {Suter}\ and\ \citenamefont
  {Jelezko}(2017)}]{suter17}%
  \BibitemOpen
  \bibfield  {author} {\bibinfo {author} {\bibfnamefont {D.}~\bibnamefont
  {Suter}}\ and\ \bibinfo {author} {\bibfnamefont {F.}~\bibnamefont
  {Jelezko}},\ }\bibfield  {title} {\enquote {\bibinfo {title} {Single-spin
  magnetic resonance in the nitrogen-vacancy center of diamond},}\ }\href@noop
  {} {\bibfield  {journal} {\bibinfo  {journal} {Prog. Nucl. Magn. Reson.
  Spectrosc.}\ }\textbf {\bibinfo {volume} {98-99}},\ \bibinfo {pages} {50--62}
  (\bibinfo {year} {2017})}\BibitemShut {NoStop}%
\bibitem [{\citenamefont {Radcliffe}(1971)}]{radcliffe71}%
  \BibitemOpen
  \bibfield  {author} {\bibinfo {author} {\bibfnamefont {J.~M.}\ \bibnamefont
  {Radcliffe}},\ }\bibfield  {title} {\enquote {\bibinfo {title} {Some
  properties of coherent spin states},}\ }\href
  {http://stacks.iop.org/0022-3689/4/i=3/a=009} {\bibfield  {journal} {\bibinfo
   {journal} {J. Phy. A}\ }\textbf {\bibinfo {volume} {4}},\ \bibinfo {pages}
  {313} (\bibinfo {year} {1971})}\BibitemShut {NoStop}%
\bibitem [{\citenamefont {Kitagawa}\ and\ \citenamefont
  {Ueda}(1993)}]{kitagawa93}%
  \BibitemOpen
  \bibfield  {author} {\bibinfo {author} {\bibfnamefont {M.}~\bibnamefont
  {Kitagawa}}\ and\ \bibinfo {author} {\bibfnamefont {M.}~\bibnamefont
  {Ueda}},\ }\bibfield  {title} {\enquote {\bibinfo {title} {Squeezed spin
  states},}\ }\href {\doibase 10.1103/PhysRevA.47.5138} {\bibfield  {journal}
  {\bibinfo  {journal} {Phys. Rev. A}\ }\textbf {\bibinfo {volume} {47}},\
  \bibinfo {pages} {5138--5143} (\bibinfo {year} {1993})}\BibitemShut {NoStop}%
\bibitem [{\citenamefont {Ma}\ \emph {et~al.}(2011)\citenamefont {Ma},
  \citenamefont {Wang}, \citenamefont {Sun},\ and\ \citenamefont
  {Nori}}]{ma11}%
  \BibitemOpen
  \bibfield  {author} {\bibinfo {author} {\bibfnamefont {J.}~\bibnamefont
  {Ma}}, \bibinfo {author} {\bibfnamefont {X.}~\bibnamefont {Wang}}, \bibinfo
  {author} {\bibfnamefont {C.~P.}\ \bibnamefont {Sun}}, \ and\ \bibinfo
  {author} {\bibfnamefont {F.}~\bibnamefont {Nori}},\ }\bibfield  {title}
  {\enquote {\bibinfo {title} {Quantum spin squeezing},}\ }\href {\doibase
  http://dx.doi.org/10.1016/j.physrep.2011.08.003} {\bibfield  {journal}
  {\bibinfo  {journal} {Phys. Rep.}\ }\textbf {\bibinfo {volume} {509}},\
  \bibinfo {pages} {89 -- 165} (\bibinfo {year} {2011})}\BibitemShut {NoStop}%
\bibitem [{\citenamefont {Dodonov}\ \emph {et~al.}(1974)\citenamefont
  {Dodonov}, \citenamefont {Malkin},\ and\ \citenamefont {Man'Ko}}]{dodonov74}%
  \BibitemOpen
  \bibfield  {author} {\bibinfo {author} {\bibfnamefont {V.~V.}\ \bibnamefont
  {Dodonov}}, \bibinfo {author} {\bibfnamefont {I.~A.}\ \bibnamefont {Malkin}},
  \ and\ \bibinfo {author} {\bibfnamefont {V.~I.}\ \bibnamefont {Man'Ko}},\
  }\bibfield  {title} {\enquote {\bibinfo {title} {Even and odd coherent states
  and excitations of a singular oscillator},}\ }\href@noop {} {\bibfield
  {journal} {\bibinfo  {journal} {Physica (Utrecht)}\ }\textbf {\bibinfo
  {volume} {72}},\ \bibinfo {pages} {597--615} (\bibinfo {year}
  {1974})}\BibitemShut {NoStop}%
\bibitem [{\citenamefont {Yurke}\ and\ \citenamefont {Stoler}(1986)}]{yurke86}%
  \BibitemOpen
  \bibfield  {author} {\bibinfo {author} {\bibfnamefont {B.}~\bibnamefont
  {Yurke}}\ and\ \bibinfo {author} {\bibfnamefont {D.}~\bibnamefont {Stoler}},\
  }\bibfield  {title} {\enquote {\bibinfo {title} {Generating quantum
  mechanical superpositions of macroscopically distinguishable states via
  amplitude dispersion},}\ }\href {\doibase 10.1103/PhysRevLett.57.13}
  {\bibfield  {journal} {\bibinfo  {journal} {Phys. Rev. Lett.}\ }\textbf
  {\bibinfo {volume} {57}},\ \bibinfo {pages} {13--16} (\bibinfo {year}
  {1986})}\BibitemShut {NoStop}%
\bibitem [{\citenamefont {Opatrn\'y}\ and\ \citenamefont
  {M\o{}lmer}(2012)}]{opatrny12}%
  \BibitemOpen
  \bibfield  {author} {\bibinfo {author} {\bibfnamefont {T.}~\bibnamefont
  {Opatrn\'y}}\ and\ \bibinfo {author} {\bibfnamefont {K.}~\bibnamefont
  {M\o{}lmer}},\ }\bibfield  {title} {\enquote {\bibinfo {title} {Spin
  squeezing and schr\"odinger-cat-state generation in atomic samples with
  rydberg blockade},}\ }\href {\doibase 10.1103/PhysRevA.86.023845} {\bibfield
  {journal} {\bibinfo  {journal} {Phys. Rev. A}\ }\textbf {\bibinfo {volume}
  {86}},\ \bibinfo {pages} {023845} (\bibinfo {year} {2012})}\BibitemShut
  {NoStop}%
\bibitem [{\citenamefont {Lau}\ \emph {et~al.}(2014)\citenamefont {Lau},
  \citenamefont {Dutton}, \citenamefont {Wang},\ and\ \citenamefont
  {Simon}}]{lau14}%
  \BibitemOpen
  \bibfield  {author} {\bibinfo {author} {\bibfnamefont {H.~W.}\ \bibnamefont
  {Lau}}, \bibinfo {author} {\bibfnamefont {Z.}~\bibnamefont {Dutton}},
  \bibinfo {author} {\bibfnamefont {T.}~\bibnamefont {Wang}}, \ and\ \bibinfo
  {author} {\bibfnamefont {C.}~\bibnamefont {Simon}},\ }\bibfield  {title}
  {\enquote {\bibinfo {title} {Proposal for the creation and optical detection
  of spin cat states in bose-einstein condensates},}\ }\href {\doibase
  10.1103/PhysRevLett.113.090401} {\bibfield  {journal} {\bibinfo  {journal}
  {Phys. Rev. Lett.}\ }\textbf {\bibinfo {volume} {113}},\ \bibinfo {pages}
  {090401} (\bibinfo {year} {2014})}\BibitemShut {NoStop}%
\bibitem [{\citenamefont {Dooley}\ \emph {et~al.}(2013)\citenamefont {Dooley},
  \citenamefont {McCrossan}, \citenamefont {Harland}, \citenamefont {Everitt},\
  and\ \citenamefont {Spiller}}]{dooley13}%
  \BibitemOpen
  \bibfield  {author} {\bibinfo {author} {\bibfnamefont {S.}~\bibnamefont
  {Dooley}}, \bibinfo {author} {\bibfnamefont {F.}~\bibnamefont {McCrossan}},
  \bibinfo {author} {\bibfnamefont {D.}~\bibnamefont {Harland}}, \bibinfo
  {author} {\bibfnamefont {M.~J.}\ \bibnamefont {Everitt}}, \ and\ \bibinfo
  {author} {\bibfnamefont {T.~P.}\ \bibnamefont {Spiller}},\ }\bibfield
  {title} {\enquote {\bibinfo {title} {Collapse and revival and cat states with
  an $n$-spin system},}\ }\href {\doibase 10.1103/PhysRevA.87.052323}
  {\bibfield  {journal} {\bibinfo  {journal} {Phys. Rev. A}\ }\textbf {\bibinfo
  {volume} {87}},\ \bibinfo {pages} {052323} (\bibinfo {year}
  {2013})}\BibitemShut {NoStop}%
\bibitem [{\citenamefont {Dooley}\ and\ \citenamefont
  {Spiller}(2014)}]{dooley14}%
  \BibitemOpen
  \bibfield  {author} {\bibinfo {author} {\bibfnamefont {S.}~\bibnamefont
  {Dooley}}\ and\ \bibinfo {author} {\bibfnamefont {T.~P.}\ \bibnamefont
  {Spiller}},\ }\bibfield  {title} {\enquote {\bibinfo {title} {Fractional
  revivals, multiple-schr\"odinger-cat states, and quantum carpets in the
  interaction of a qubit with $n$ qubits},}\ }\href {\doibase
  10.1103/PhysRevA.90.012320} {\bibfield  {journal} {\bibinfo  {journal} {Phys.
  Rev. A}\ }\textbf {\bibinfo {volume} {90}},\ \bibinfo {pages} {012320}
  (\bibinfo {year} {2014})}\BibitemShut {NoStop}%
\bibitem [{\citenamefont {Huang}\ \emph {et~al.}(2015)\citenamefont {Huang},
  \citenamefont {Qin}, \citenamefont {Zhong}, \citenamefont {Ke},\ and\
  \citenamefont {Lee}}]{huang15}%
  \BibitemOpen
  \bibfield  {author} {\bibinfo {author} {\bibfnamefont {J.}~\bibnamefont
  {Huang}}, \bibinfo {author} {\bibfnamefont {X.}~\bibnamefont {Qin}}, \bibinfo
  {author} {\bibfnamefont {H.}~\bibnamefont {Zhong}}, \bibinfo {author}
  {\bibfnamefont {Y.}~\bibnamefont {Ke}}, \ and\ \bibinfo {author}
  {\bibfnamefont {C.}~\bibnamefont {Lee}},\ }\bibfield  {title} {\enquote
  {\bibinfo {title} {Quantum metrology with spin cat states under
  dissipation},}\ }\href@noop {} {\bibfield  {journal} {\bibinfo  {journal}
  {Sci. Rep.}\ }\textbf {\bibinfo {volume} {5}},\ \bibinfo {pages} {17894}
  (\bibinfo {year}{2015})}\BibitemShut {NoStop}%
\bibitem [{\citenamefont {Pezz{\`e}}\ \emph {et~al.}(2016)\citenamefont
  {Pezz{\`e}}, \citenamefont {Smerzi}, \citenamefont {Oberthaler},
  \citenamefont {Schmied},\ and\ \citenamefont {Treutlein}}]{pezze16}%
  \BibitemOpen
  \bibfield  {author} {\bibinfo {author} {\bibfnamefont {L.}~\bibnamefont
  {Pezz{\`e}}}, \bibinfo {author} {\bibfnamefont {A.}~\bibnamefont {Smerzi}},
  \bibinfo {author} {\bibfnamefont {M.~K.}\ \bibnamefont {Oberthaler}},
  \bibinfo {author} {\bibfnamefont {R.}~\bibnamefont {Schmied}}, \ and\
  \bibinfo {author} {\bibfnamefont {P.}~\bibnamefont {Treutlein}},\ }\bibfield
  {title} {\enquote {\bibinfo {title} {Non-classical states of atomic
  ensembles: fundamentals and applications in quantum metrology},}\ }\href@noop
  {} {\bibfield  {journal} {\bibinfo  {journal} {arXiv preprint
  arXiv:1609.01609}\ } (\bibinfo {year} {2016})}\BibitemShut {NoStop}%
\bibitem [{\citenamefont {Cohen}\ and\ \citenamefont {Reif}(1957)}]{cohen57}%
  \BibitemOpen
  \bibfield  {author} {\bibinfo {author} {\bibfnamefont {M.~H.}\ \bibnamefont
  {Cohen}}\ and\ \bibinfo {author} {\bibfnamefont {F.}~\bibnamefont {Reif}},\
  }\bibfield  {title} {\enquote {\bibinfo {title} {Quadrupole effects in
  nuclear magnetic resonance studies of solids},}\ }\href@noop {} {\bibfield
  {journal} {\bibinfo  {journal} {Solid State Phys.}\ }\textbf {\bibinfo
  {volume} {5}},\ \bibinfo {pages} {321} (\bibinfo {year} {1957})}\BibitemShut
  {NoStop}%
\bibitem [{\citenamefont {Das}\ and\ \citenamefont {Hahn}(1958)}]{das58}%
  \BibitemOpen
  \bibfield  {author} {\bibinfo {author} {\bibfnamefont {T.~P.}\ \bibnamefont
  {Das}}\ and\ \bibinfo {author} {\bibfnamefont {E.~L.}\ \bibnamefont {Hahn}},\
  }\href@noop {} {\emph {\bibinfo {title} {Nuclear Quadrupole Resonance
  Spectroscopy}}}\ (\bibinfo  {publisher} {Academic Press},\ \bibinfo {address}
  {New York},\ \bibinfo {year} {1958})\BibitemShut {NoStop}%
\bibitem [{\citenamefont {Auccaise}\ \emph {et~al.}(2015)\citenamefont
  {Auccaise}, \citenamefont {Araujo-Ferreira}, \citenamefont {Sarthour},
  \citenamefont {Oliveira}, \citenamefont {Bonagamba},\ and\ \citenamefont
  {Roditi}}]{auccaise15}%
  \BibitemOpen
  \bibfield  {author} {\bibinfo {author} {\bibfnamefont {R.}~\bibnamefont
  {Auccaise}}, \bibinfo {author} {\bibfnamefont {A.~G.}\ \bibnamefont
  {Araujo-Ferreira}}, \bibinfo {author} {\bibfnamefont {R.~S.}\ \bibnamefont
  {Sarthour}}, \bibinfo {author} {\bibfnamefont {I.~S.}\ \bibnamefont
  {Oliveira}}, \bibinfo {author} {\bibfnamefont {T.~J.}\ \bibnamefont
  {Bonagamba}}, \ and\ \bibinfo {author} {\bibfnamefont {I.}~\bibnamefont
  {Roditi}},\ }\bibfield  {title} {\enquote {\bibinfo {title} {Spin squeezing
  in a quadrupolar nuclei nmr system},}\ }\href {\doibase
  10.1103/PhysRevLett.114.043604} {\bibfield  {journal} {\bibinfo  {journal}
  {Phys. Rev. Lett.}\ }\textbf {\bibinfo {volume} {114}},\ \bibinfo {pages}
  {043604} (\bibinfo {year} {2015})}\BibitemShut {NoStop}%
\bibitem [{\citenamefont {Aksu~Korkmaz}\ and\ \citenamefont
  {Bulutay}(2016)}]{korkmaz16}%
  \BibitemOpen
  \bibfield  {author} {\bibinfo {author} {\bibfnamefont {Y.}~\bibnamefont
  {Aksu~Korkmaz}}\ and\ \bibinfo {author} {\bibfnamefont {C.}~\bibnamefont
  {Bulutay}},\ }\bibfield  {title} {\enquote {\bibinfo {title} {Nuclear spin
  squeezing via electric quadrupole interaction},}\ }\href {\doibase
  10.1103/PhysRevA.93.013812} {\bibfield  {journal} {\bibinfo  {journal} {Phys.
  Rev. A}\ }\textbf {\bibinfo {volume} {93}},\ \bibinfo {pages} {013812}
  (\bibinfo {year} {2016})}\BibitemShut {NoStop}%
\bibitem [{\citenamefont {Jin}\ and\ \citenamefont {Kim}(2007)}]{jin07}%
  \BibitemOpen
  \bibfield  {author} {\bibinfo {author} {\bibfnamefont {G.-R.}\ \bibnamefont
  {Jin}}\ and\ \bibinfo {author} {\bibfnamefont {S.~W.}\ \bibnamefont {Kim}},\
  }\bibfield  {title} {\enquote {\bibinfo {title} {Storage of spin squeezing in
  a two-component bose-einstein condensate},}\ }\href {\doibase
  10.1103/PhysRevLett.99.170405} {\bibfield  {journal} {\bibinfo  {journal}
  {Phys. Rev. Lett.}\ }\textbf {\bibinfo {volume} {99}},\ \bibinfo {pages}
  {170405} (\bibinfo {year} {2007})}\BibitemShut {NoStop}%
\bibitem [{\citenamefont {Wu}\ \emph {et~al.}(2015)\citenamefont {Wu},
  \citenamefont {Tey},\ and\ \citenamefont {You}}]{wu15}%
  \BibitemOpen
  \bibfield  {author} {\bibinfo {author} {\bibfnamefont {L.-N.}\ \bibnamefont
  {Wu}}, \bibinfo {author} {\bibfnamefont {M.~K.}\ \bibnamefont {Tey}}, \ and\
  \bibinfo {author} {\bibfnamefont {L.}~\bibnamefont {You}},\ }\bibfield
  {title} {\enquote {\bibinfo {title} {Persistent atomic spin squeezing at the
  heisenberg limit},}\ }\href {\doibase 10.1103/PhysRevA.92.063610} {\bibfield
  {journal} {\bibinfo  {journal} {Phys. Rev. A}\ }\textbf {\bibinfo {volume}
  {92}},\ \bibinfo {pages} {063610} (\bibinfo {year} {2015})}\BibitemShut
  {NoStop}%
\bibitem [{\citenamefont {Kajtoch}\ \emph {et~al.}(2016)\citenamefont
  {Kajtoch}, \citenamefont {Paw\l{}owski},\ and\ \citenamefont
  {Witkowska}}]{kajtoch16}%
  \BibitemOpen
  \bibfield  {author} {\bibinfo {author} {\bibfnamefont {D.}~\bibnamefont
  {Kajtoch}}, \bibinfo {author} {\bibfnamefont {K.}~\bibnamefont
  {Paw\l{}owski}}, \ and\ \bibinfo {author} {\bibfnamefont {E.}~\bibnamefont
  {Witkowska}},\ }\bibfield  {title} {\enquote {\bibinfo {title} {Entanglement
  storage by classical fixed points in the two-axis countertwisting model},}\
  }\href {\doibase 10.1103/PhysRevA.93.022331} {\bibfield  {journal} {\bibinfo
  {journal} {Phys. Rev. A}\ }\textbf {\bibinfo {volume} {93}},\ \bibinfo
  {pages} {022331} (\bibinfo {year} {2016})}\BibitemShut {NoStop}%
\bibitem [{\citenamefont {Cummins}\ \emph {et~al.}(2003)\citenamefont
  {Cummins}, \citenamefont {Llewellyn},\ and\ \citenamefont
  {Jones}}]{cummins03}%
  \BibitemOpen
  \bibfield  {author} {\bibinfo {author} {\bibfnamefont {H.~K.}\ \bibnamefont
  {Cummins}}, \bibinfo {author} {\bibfnamefont {G.}~\bibnamefont {Llewellyn}},
  \ and\ \bibinfo {author} {\bibfnamefont {J.~A.}\ \bibnamefont {Jones}},\
  }\bibfield  {title} {\enquote {\bibinfo {title} {Tackling systematic errors
  in quantum logic gates with composite rotations},}\ }\href {\doibase
  10.1103/PhysRevA.67.042308} {\bibfield  {journal} {\bibinfo  {journal} {Phys.
  Rev. A}\ }\textbf {\bibinfo {volume} {67}},\ \bibinfo {pages} {042308}
  (\bibinfo {year} {2003})}\BibitemShut {NoStop}%
\bibitem [{\citenamefont {Mirrahimi}\ \emph {et~al.}(2014)\citenamefont
  {Mirrahimi}, \citenamefont {Leghtas}, \citenamefont {Albert}, \citenamefont
  {Touzard}, \citenamefont {Schoelkopf}, \citenamefont {Jiang},\ and\
  \citenamefont {Devoret}}]{mirrahimi14}%
  \BibitemOpen
  \bibfield  {author} {\bibinfo {author} {\bibfnamefont {M.}~\bibnamefont
  {Mirrahimi}}, \bibinfo {author} {\bibfnamefont {Z.}~\bibnamefont {Leghtas}},
  \bibinfo {author} {\bibfnamefont {V.~V.}\ \bibnamefont {Albert}}, \bibinfo
  {author} {\bibfnamefont {S.}~\bibnamefont {Touzard}}, \bibinfo {author}
  {\bibfnamefont {R.~J.}\ \bibnamefont {Schoelkopf}}, \bibinfo {author}
  {\bibfnamefont {L.}~\bibnamefont {Jiang}}, \ and\ \bibinfo {author}
  {\bibfnamefont {M.~H.}\ \bibnamefont {Devoret}},\ }\bibfield  {title}
  {\enquote {\bibinfo {title} {Dynamically protected cat-qubits: a new paradigm
  for universal quantum computation},}\ }\href
  {http://stacks.iop.org/1367-2630/16/i=4/a=045014} {\bibfield  {journal}
  {\bibinfo  {journal} {New J. Phys.}\ }\textbf {\bibinfo {volume} {16}},\
  \bibinfo {pages} {045014} (\bibinfo {year} {2014})}\BibitemShut {NoStop}%
\bibitem [{\citenamefont {Albert}\ \emph {et~al.}(2016)\citenamefont {Albert},
  \citenamefont {Shu}, \citenamefont {Krastanov}, \citenamefont {Shen},
  \citenamefont {Liu}, \citenamefont {Yang}, \citenamefont {Schoelkopf},
  \citenamefont {Mirrahimi}, \citenamefont {Devoret},\ and\ \citenamefont
  {Jiang}}]{albert16}%
  \BibitemOpen
  \bibfield  {author} {\bibinfo {author} {\bibfnamefont {V.~V.}\ \bibnamefont
  {Albert}}, \bibinfo {author} {\bibfnamefont {C.}~\bibnamefont {Shu}},
  \bibinfo {author} {\bibfnamefont {S.}~\bibnamefont {Krastanov}}, \bibinfo
  {author} {\bibfnamefont {C.}~\bibnamefont {Shen}}, \bibinfo {author}
  {\bibfnamefont {R.-B.}\ \bibnamefont {Liu}}, \bibinfo {author} {\bibfnamefont
  {Z.-B.}\ \bibnamefont {Yang}}, \bibinfo {author} {\bibfnamefont {R.~J.}\
  \bibnamefont {Schoelkopf}}, \bibinfo {author} {\bibfnamefont
  {M.}~\bibnamefont {Mirrahimi}}, \bibinfo {author} {\bibfnamefont {M.~H.}\
  \bibnamefont {Devoret}}, \ and\ \bibinfo {author} {\bibfnamefont
  {L.}~\bibnamefont {Jiang}},\ }\bibfield  {title} {\enquote {\bibinfo {title}
  {Holonomic quantum control with continuous variable systems},}\ }\href
  {\doibase 10.1103/PhysRevLett.116.140502} {\bibfield  {journal} {\bibinfo
  {journal} {Phys. Rev. Lett.}\ }\textbf {\bibinfo {volume} {116}},\ \bibinfo
  {pages} {140502} (\bibinfo {year} {2016})}\BibitemShut {NoStop}%
\bibitem [{\citenamefont {Bulutay}(2012)}]{bulutay12}%
  \BibitemOpen
  \bibfield  {author} {\bibinfo {author} {\bibfnamefont {C.}~\bibnamefont
  {Bulutay}},\ }\bibfield  {title} {\enquote {\bibinfo {title} {Quadrupolar
  spectra of nuclear spins in strained ${\rm in}_{x}{\rm
  ga}_{1\ensuremath{-}x}{\rm as}$ quantum dots},}\ }\href {\doibase
  10.1103/PhysRevB.85.115313} {\bibfield  {journal} {\bibinfo  {journal} {Phys.
  Rev. B}\ }\textbf {\bibinfo {volume} {85}},\ \bibinfo {pages} {115313}
  (\bibinfo {year} {2012})}\BibitemShut {NoStop}%
\bibitem [{\citenamefont {Sanders}(1989)}]{sanders89}%
  \BibitemOpen
  \bibfield  {author} {\bibinfo {author} {\bibfnamefont {B.~C.}\ \bibnamefont
  {Sanders}},\ }\bibfield  {title} {\enquote {\bibinfo {title} {Quantum
  dynamics of the nonlinear rotator and the effects of continual spin
  measurement},}\ }\href {\doibase 10.1103/PhysRevA.40.2417} {\bibfield
  {journal} {\bibinfo  {journal} {Phys. Rev. A}\ }\textbf {\bibinfo {volume}
  {40}},\ \bibinfo {pages} {2417--2427} (\bibinfo {year} {1989})}\BibitemShut
  {NoStop}%
\bibitem [{\citenamefont {Gilchrist}\ \emph {et~al.}(2004)\citenamefont
  {Gilchrist}, \citenamefont {Nemoto}, \citenamefont {Munro}, \citenamefont
  {Ralph}, \citenamefont {Glancy}, \citenamefont {Braunstein},\ and\
  \citenamefont {Milburn}}]{gilchrist04}%
  \BibitemOpen
  \bibfield  {author} {\bibinfo {author} {\bibfnamefont {A.}~\bibnamefont
  {Gilchrist}}, \bibinfo {author} {\bibfnamefont {K.}~\bibnamefont {Nemoto}},
  \bibinfo {author} {\bibfnamefont {W.~J.}\ \bibnamefont {Munro}}, \bibinfo
  {author} {\bibfnamefont {T.~C.}\ \bibnamefont {Ralph}}, \bibinfo {author}
  {\bibfnamefont {S.}~\bibnamefont {Glancy}}, \bibinfo {author} {\bibfnamefont
  {S.~L.}\ \bibnamefont {Braunstein}}, \ and\ \bibinfo {author} {\bibfnamefont
  {G.~J.}\ \bibnamefont {Milburn}},\ }\bibfield  {title} {\enquote {\bibinfo
  {title} {Schr\"{o}dinger cats and their power for quantum information
  processing},}\ }\href {http://stacks.iop.org/1464-4266/6/i=8/a=032}
  {\bibfield  {journal} {\bibinfo  {journal} {J. Opt. B Quantum Semiclassical
  Opt.}\ }\textbf {\bibinfo {volume} {6}},\ \bibinfo {pages} {S828} (\bibinfo
  {year} {2004})}\BibitemShut {NoStop}%
\bibitem [{\citenamefont {Gerry}\ and\ \citenamefont
  {Knight}(2005)}]{gerry-knight}%
  \BibitemOpen
  \bibfield  {author} {\bibinfo {author} {\bibfnamefont {C.~C.}\ \bibnamefont
  {Gerry}}\ and\ \bibinfo {author} {\bibfnamefont {P.~L.}\ \bibnamefont
  {Knight}},\ }\href@noop {} {\emph {\bibinfo {title} {Introductory quantum
  optics}}}\ (\bibinfo  {publisher} {Cambridge University Press},\ \bibinfo
  {year} {2005})\BibitemShut {NoStop}%
\bibitem [{\citenamefont {Nielsen}\ and\ \citenamefont
  {Chuang}(2000)}]{nielsen-book}%
  \BibitemOpen
  \bibfield  {author} {\bibinfo {author} {\bibfnamefont {M.~A.}\ \bibnamefont
  {Nielsen}}\ and\ \bibinfo {author} {\bibfnamefont {I.~L.}\ \bibnamefont
  {Chuang}},\ }\href@noop {} {\emph {\bibinfo {title} {Quantum Computation and
  Quantum Information}}}\ (\bibinfo  {publisher} {Cambridge University Press},\
  \bibinfo {address} {New York},\ \bibinfo {year} {2000})\BibitemShut {NoStop}%
\bibitem [{\citenamefont {Bj\"{o}rk}\ and\ \citenamefont
  {Mana}(2004)}]{bjork04}%
  \BibitemOpen
  \bibfield  {author} {\bibinfo {author} {\bibfnamefont {G.}~\bibnamefont
  {Bj\"{o}rk}}\ and\ \bibinfo {author} {\bibfnamefont {P.~G.~Luca}\
  \bibnamefont {Mana}},\ }\bibfield  {title} {\enquote {\bibinfo {title} {A
  size criterion for macroscopic superposition states},}\ }\href
  {http://stacks.iop.org/1464-4266/6/i=11/a=001} {\bibfield  {journal}
  {\bibinfo  {journal} {J. Opt. B Quantum Semiclassical Opt.}\ }\textbf
  {\bibinfo {volume} {6}},\ \bibinfo {pages} {429} (\bibinfo {year}
  {2004})}\BibitemShut {NoStop}%
\bibitem [{\citenamefont {Fr{\"o}wis}\ and\ \citenamefont
  {D{\"u}r}(2012)}]{frowis12}%
  \BibitemOpen
  \bibfield  {author} {\bibinfo {author} {\bibfnamefont {F.}~\bibnamefont
  {Fr{\"o}wis}}\ and\ \bibinfo {author} {\bibfnamefont {W.}~\bibnamefont
  {D{\"u}r}},\ }\bibfield  {title} {\enquote {\bibinfo {title} {Measures of
  macroscopicity for quantum spin systems},}\ }\href@noop {} {\bibfield
  {journal} {\bibinfo  {journal} {New J. Phys.}\ }\textbf {\bibinfo {volume}
  {14}},\ \bibinfo {pages} {093039} (\bibinfo {year} {2012})}\BibitemShut
  {NoStop}%
\bibitem [{\citenamefont {Pirandola}\ \emph {et~al.}(2008)\citenamefont
  {Pirandola}, \citenamefont {Mancini}, \citenamefont {Braunstein},\ and\
  \citenamefont {Vitali}}]{pirandola08}%
  \BibitemOpen
  \bibfield  {author} {\bibinfo {author} {\bibfnamefont {S.}~\bibnamefont
  {Pirandola}}, \bibinfo {author} {\bibfnamefont {S.}~\bibnamefont {Mancini}},
  \bibinfo {author} {\bibfnamefont {S.~L.}\ \bibnamefont {Braunstein}}, \ and\
  \bibinfo {author} {\bibfnamefont {D.}~\bibnamefont {Vitali}},\ }\bibfield
  {title} {\enquote {\bibinfo {title} {Minimal qudit code for a qubit in the
  phase-damping channel},}\ }\href {\doibase 10.1103/PhysRevA.77.032309}
  {\bibfield  {journal} {\bibinfo  {journal} {Phys. Rev. A}\ }\textbf {\bibinfo
  {volume} {77}},\ \bibinfo {pages} {032309} (\bibinfo {year}
  {2008})}\BibitemShut {NoStop}%
\bibitem [{\citenamefont {Huang}\ \emph {et~al.}(2012)\citenamefont {Huang},
  \citenamefont {Ma}, \citenamefont {Jing},\ and\ \citenamefont
  {Wang}}]{huang12}%
  \BibitemOpen
  \bibfield  {author} {\bibinfo {author} {\bibfnamefont {Y.-X.}\ \bibnamefont
  {Huang}}, \bibinfo {author} {\bibfnamefont {J.}~\bibnamefont {Ma}}, \bibinfo
  {author} {\bibfnamefont {X.-X.}\ \bibnamefont {Jing}}, \ and\ \bibinfo
  {author} {\bibfnamefont {X.-G.}\ \bibnamefont {Wang}},\ }\bibfield  {title}
  {\enquote {\bibinfo {title} {Spin squeezing and fixed-point bifurcation},}\
  }\href {http://stacks.iop.org/0253-6102/58/i=6/a=03} {\bibfield  {journal}
  {\bibinfo  {journal} {Commun. Theor. Phys.}\ }\textbf {\bibinfo {volume}
  {58}},\ \bibinfo {pages} {800} (\bibinfo {year} {2012})}\BibitemShut
  {NoStop}%
\bibitem [{\citenamefont {Johansson}\ \emph {et~al.}(2012)\citenamefont
  {Johansson}, \citenamefont {Nation},\ and\ \citenamefont {Nori}}]{qutip1}%
  \BibitemOpen
  \bibfield  {author} {\bibinfo {author} {\bibfnamefont {J.~R.}\ \bibnamefont
  {Johansson}}, \bibinfo {author} {\bibfnamefont {P.~D.}\ \bibnamefont
  {Nation}}, \ and\ \bibinfo {author} {\bibfnamefont {Franco}\ \bibnamefont
  {Nori}},\ }\bibfield  {title} {\enquote {\bibinfo {title} {Qutip: An
  open-source python framework for the dynamics of open quantum systems},}\
  }\href@noop {} {\bibfield  {journal} {\bibinfo  {journal} {Comput. Phys.
  Commun.}\ }\textbf {\bibinfo {volume} {183}},\ \bibinfo {pages} {1760--1772}
  (\bibinfo {year} {2012})}\BibitemShut {NoStop}%
\bibitem [{\citenamefont {Johansson}\ \emph {et~al.}(2013)\citenamefont
  {Johansson}, \citenamefont {Nation},\ and\ \citenamefont {Nori}}]{qutip2}%
  \BibitemOpen
  \bibfield  {author} {\bibinfo {author} {\bibfnamefont {J.~R.}\ \bibnamefont
  {Johansson}}, \bibinfo {author} {\bibfnamefont {P.~D.}\ \bibnamefont
  {Nation}}, \ and\ \bibinfo {author} {\bibfnamefont {Franco}\ \bibnamefont
  {Nori}},\ }\bibfield  {title} {\enquote {\bibinfo {title} {Qutip2:a python
  framework for the dynamics of open quantum systems},}\ }\href@noop {}
  {\bibfield  {journal} {\bibinfo  {journal} {Comput. Phys. Commun.}\ }\textbf
  {\bibinfo {volume} {184}},\ \bibinfo {pages} {1234--1240} (\bibinfo {year}
  {2013})}\BibitemShut {NoStop}%
\bibitem [{\citenamefont {Agarwal}\ \emph {et~al.}(1997)\citenamefont
  {Agarwal}, \citenamefont {Puri},\ and\ \citenamefont {Singh}}]{agarwal97}%
  \BibitemOpen
  \bibfield  {author} {\bibinfo {author} {\bibfnamefont {G.~S.}\ \bibnamefont
  {Agarwal}}, \bibinfo {author} {\bibfnamefont {R.~R.}\ \bibnamefont {Puri}}, \
  and\ \bibinfo {author} {\bibfnamefont {R.~P.}\ \bibnamefont {Singh}},\
  }\bibfield  {title} {\enquote {\bibinfo {title} {Atomic schr\"odinger cat
  states},}\ }\href {\doibase 10.1103/PhysRevA.56.2249} {\bibfield  {journal}
  {\bibinfo  {journal} {Phys. Rev. A}\ }\textbf {\bibinfo {volume} {56}},\
  \bibinfo {pages} {2249--2254} (\bibinfo {year} {1997})}\BibitemShut {NoStop}%
\bibitem [{\citenamefont {Chumakov}\ \emph {et~al.}(1999)\citenamefont
  {Chumakov}, \citenamefont {Frank},\ and\ \citenamefont {Wolf}}]{chumakov99}%
  \BibitemOpen
  \bibfield  {author} {\bibinfo {author} {\bibfnamefont {S.~M.}\ \bibnamefont
  {Chumakov}}, \bibinfo {author} {\bibfnamefont {A.}~\bibnamefont {Frank}}, \
  and\ \bibinfo {author} {\bibfnamefont {K.~B.}\ \bibnamefont {Wolf}},\
  }\bibfield  {title} {\enquote {\bibinfo {title} {Finite kerr medium:
  Macroscopic quantum superposition states and wigner functions on the
  sphere},}\ }\href {\doibase 10.1103/PhysRevA.60.1817} {\bibfield  {journal}
  {\bibinfo  {journal} {Phys. Rev. A}\ }\textbf {\bibinfo {volume} {60}},\
  \bibinfo {pages} {1817--1823} (\bibinfo {year} {1999})}\BibitemShut {NoStop}%
\bibitem [{\citenamefont {Roy}\ \emph {et~al.}(2016)\citenamefont {Roy},
  \citenamefont {Stone},\ and\ \citenamefont {Jiang}}]{roy16}%
  \BibitemOpen
  \bibfield  {author} {\bibinfo {author} {\bibfnamefont {A.}~\bibnamefont
  {Roy}}, \bibinfo {author} {\bibfnamefont {A.~D.}\ \bibnamefont {Stone}}, \
  and\ \bibinfo {author} {\bibfnamefont {L.}~\bibnamefont {Jiang}},\ }\bibfield
   {title} {\enquote {\bibinfo {title} {Concurrent remote entanglement with
  quantum error correction against photon losses},}\ }\href {\doibase
  10.1103/PhysRevA.94.032333} {\bibfield  {journal} {\bibinfo  {journal} {Phys.
  Rev. A}\ }\textbf {\bibinfo {volume} {94}},\ \bibinfo {pages} {032333}
  (\bibinfo {year} {2016})}\BibitemShut {NoStop}%
\bibitem [{\citenamefont {Bhattacharya}(2015)}]{bhattacharya15}%
  \BibitemOpen
  \bibfield  {author} {\bibinfo {author} {\bibfnamefont {M.}~\bibnamefont
  {Bhattacharya}},\ }\bibfield  {title} {\enquote {\bibinfo {title} {Analytical
  solvability of the two-axis countertwisting spin squeezing hamiltonian},}\
  }\href@noop {} {\bibfield  {journal} {\bibinfo  {journal} {arXiv preprint
  arXiv:1509.08530}\ } (\bibinfo {year} {2015})}\BibitemShut {NoStop}%
\bibitem [{\citenamefont {Brune}\ \emph {et~al.}(1996)\citenamefont {Brune},
  \citenamefont {Hagley}, \citenamefont {Dreyer}, \citenamefont {Ma\^{\i}tre},
  \citenamefont {Maali}, \citenamefont {Wunderlich}, \citenamefont {Raimond},\
  and\ \citenamefont {Haroche}}]{brune96}%
  \BibitemOpen
  \bibfield  {author} {\bibinfo {author} {\bibfnamefont {M.}~\bibnamefont
  {Brune}}, \bibinfo {author} {\bibfnamefont {E.}~\bibnamefont {Hagley}},
  \bibinfo {author} {\bibfnamefont {J.}~\bibnamefont {Dreyer}}, \bibinfo
  {author} {\bibfnamefont {X.}~\bibnamefont {Ma\^{\i}tre}}, \bibinfo {author}
  {\bibfnamefont {A.}~\bibnamefont {Maali}}, \bibinfo {author} {\bibfnamefont
  {C.}~\bibnamefont {Wunderlich}}, \bibinfo {author} {\bibfnamefont {J.~M.}\
  \bibnamefont {Raimond}}, \ and\ \bibinfo {author} {\bibfnamefont
  {S.}~\bibnamefont {Haroche}},\ }\bibfield  {title} {\enquote {\bibinfo
  {title} {Observing the progressive decoherence of the ``meter'' in a quantum
  measurement},}\ }\href {\doibase 10.1103/PhysRevLett.77.4887} {\bibfield
  {journal} {\bibinfo  {journal} {Phys. Rev. Lett.}\ }\textbf {\bibinfo
  {volume} {77}},\ \bibinfo {pages} {4887--4890} (\bibinfo {year}
  {1996})}\BibitemShut {NoStop}%
\bibitem [{\citenamefont {Byrnes}\ \emph {et~al.}(2012)\citenamefont {Byrnes},
  \citenamefont {Wen},\ and\ \citenamefont {Yamamoto}}]{byrnes12}%
  \BibitemOpen
  \bibfield  {author} {\bibinfo {author} {\bibfnamefont {T.}~\bibnamefont
  {Byrnes}}, \bibinfo {author} {\bibfnamefont {K.}~\bibnamefont {Wen}}, \ and\
  \bibinfo {author} {\bibfnamefont {Y.}~\bibnamefont {Yamamoto}},\ }\bibfield
  {title} {\enquote {\bibinfo {title} {Macroscopic quantum computation using
  bose-einstein condensates},}\ }\href {\doibase 10.1103/PhysRevA.85.040306}
  {\bibfield  {journal} {\bibinfo  {journal} {Phys. Rev. A}\ }\textbf {\bibinfo
  {volume} {85}},\ \bibinfo {pages} {040306} (\bibinfo {year}
  {2012})}\BibitemShut {NoStop}%
\bibitem [{\citenamefont {Haroche}\ and\ \citenamefont
  {Raimond}(2006)}]{haroche06}%
  \BibitemOpen
  \bibfield  {author} {\bibinfo {author} {\bibfnamefont {S.}~\bibnamefont
  {Haroche}}\ and\ \bibinfo {author} {\bibfnamefont {J.-M.}\ \bibnamefont
  {Raimond}},\ }\href@noop {} {\emph {\bibinfo {title} {Exploring the
  quantum}}}\ (\bibinfo  {publisher} {Oxford University Press, Oxford},\
  \bibinfo {year} {2006})\BibitemShut {NoStop}%
\bibitem [{\citenamefont {Wallraff}\ \emph {et~al.}(2004)\citenamefont
  {Wallraff}, \citenamefont {Schuster}, \citenamefont {Blais}, \citenamefont
  {Frunzio}, \citenamefont {Huang}, \citenamefont {Majer}, \citenamefont
  {Kumar}, \citenamefont {Girvin},\ and\ \citenamefont
  {Schoelkopf}}]{wallraff04}%
  \BibitemOpen
  \bibfield  {author} {\bibinfo {author} {\bibfnamefont {A.}~\bibnamefont
  {Wallraff}}, \bibinfo {author} {\bibfnamefont {D.~I.}\ \bibnamefont
  {Schuster}}, \bibinfo {author} {\bibfnamefont {A.}~\bibnamefont {Blais}},
  \bibinfo {author} {\bibfnamefont {L.}~\bibnamefont {Frunzio}}, \bibinfo
  {author} {\bibfnamefont {R.-S.}\ \bibnamefont {Huang}}, \bibinfo {author}
  {\bibfnamefont {J.}~\bibnamefont {Majer}}, \bibinfo {author} {\bibfnamefont
  {S.}~\bibnamefont {Kumar}}, \bibinfo {author} {\bibfnamefont {S.~M.}\
  \bibnamefont {Girvin}}, \ and\ \bibinfo {author} {\bibfnamefont {R.~J.}\
  \bibnamefont {Schoelkopf}},\ }\bibfield  {title} {\enquote {\bibinfo {title}
  {Strong coupling of a single photon to a superconducting qubit using circuit
  quantum electrodynamics},}\ }\href@noop {} {\bibfield  {journal} {\bibinfo
  {journal} {Nature (London)}\ }\textbf {\bibinfo {volume} {431}},\ \bibinfo
  {pages} {162--167} (\bibinfo {year} {2004})}\BibitemShut {NoStop}%
\bibitem [{\citenamefont {M{\"u}ller}\ \emph {et~al.}(2014)\citenamefont
  {M{\"u}ller}, \citenamefont {Kong}, \citenamefont {Cai}, \citenamefont
  {Melentijevi\'{c}}, \citenamefont {Stacey}, \citenamefont {Markham},
  \citenamefont {Twitchen}, \citenamefont {Isoya}, \citenamefont {Pezzagna},
  \citenamefont {Meijer}, \citenamefont {Du}, \citenamefont {Plenio},
  \citenamefont {Naydenov}, \citenamefont {McGuinness},\ and\ \citenamefont
  {Jelezko}}]{muller14}%
  \BibitemOpen
  \bibfield  {author} {\bibinfo {author} {\bibfnamefont {C.}~\bibnamefont
  {M{\"u}ller}}, \bibinfo {author} {\bibfnamefont {X.}~\bibnamefont {Kong}},
  \bibinfo {author} {\bibfnamefont {J.-M.}\ \bibnamefont {Cai}}, \bibinfo
  {author} {\bibfnamefont {K.}~\bibnamefont {Melentijevi\'{c}}}, \bibinfo
  {author} {\bibfnamefont {A.}~\bibnamefont {Stacey}}, \bibinfo {author}
  {\bibfnamefont {M.}~\bibnamefont {Markham}}, \bibinfo {author} {\bibfnamefont
  {D.}~\bibnamefont {Twitchen}}, \bibinfo {author} {\bibfnamefont
  {J.}~\bibnamefont {Isoya}}, \bibinfo {author} {\bibfnamefont
  {S.}~\bibnamefont {Pezzagna}}, \bibinfo {author} {\bibfnamefont
  {J.}~\bibnamefont {Meijer}}, \bibinfo {author} {\bibfnamefont {J.~F.}\
  \bibnamefont {Du}}, \bibinfo {author} {\bibfnamefont {M.~B.}\ \bibnamefont
  {Plenio}}, \bibinfo {author} {\bibfnamefont {B.}~\bibnamefont {Naydenov}},
  \bibinfo {author} {\bibfnamefont {L.~P.}\ \bibnamefont {McGuinness}}, \ and\
  \bibinfo {author} {\bibfnamefont {F.}~\bibnamefont {Jelezko}},\ }\bibfield
  {title} {\enquote {\bibinfo {title} {Nuclear magnetic resonance spectroscopy
  with single spin sensitivity},}\ }\href {\doibase 10.1038/ncomms5703}
  {\bibfield  {journal} {\bibinfo  {journal} {Nat. Commun.}\ }\textbf {\bibinfo
  {volume} {5}},\ \bibinfo {pages} {4703} (\bibinfo {year} {2014})}\BibitemShut
  {NoStop}%
\bibitem [{\citenamefont {W\"ust}\ \emph {et~al.}(2016)\citenamefont {W\"ust},
  \citenamefont {Munsch}, \citenamefont {Maier}, \citenamefont {Kuhlmann},
  \citenamefont {Ludwig}, \citenamefont {Wieck}, \citenamefont {Loss},
  \citenamefont {Poggio},\ and\ \citenamefont {Warburton}}]{wust16}%
  \BibitemOpen
  \bibfield  {author} {\bibinfo {author} {\bibfnamefont {G.}~\bibnamefont
  {W\"ust}}, \bibinfo {author} {\bibfnamefont {M.}~\bibnamefont {Munsch}},
  \bibinfo {author} {\bibfnamefont {F.}~\bibnamefont {Maier}}, \bibinfo
  {author} {\bibfnamefont {A.~V.}\ \bibnamefont {Kuhlmann}}, \bibinfo {author}
  {\bibfnamefont {A.}~\bibnamefont {Ludwig}}, \bibinfo {author} {\bibfnamefont
  {A.~D.}\ \bibnamefont {Wieck}}, \bibinfo {author} {\bibfnamefont
  {D.}~\bibnamefont {Loss}}, \bibinfo {author} {\bibfnamefont {M.}~\bibnamefont
  {Poggio}}, \ and\ \bibinfo {author} {\bibfnamefont {R.~J.}\ \bibnamefont
  {Warburton}},\ }\bibfield  {title} {\enquote {\bibinfo {title} {Role of the
  electron spin in determining the coherence of the nuclear spins in a quantum
  dot},}\ }\href {\doibase 10.1038/nnano.2016.114} {\bibfield  {journal}
  {\bibinfo  {journal} {Nat. Nanotechnol.}\ }\textbf {\bibinfo {volume} {11}},\
  \bibinfo {pages} {885} (\bibinfo {year} {2016})}\BibitemShut {NoStop}%
\bibitem [{\citenamefont {Chekhovich}\ \emph {et~al.}(2015)\citenamefont
  {Chekhovich}, \citenamefont {Hopkinson}, \citenamefont {Skolnick},\ and\
  \citenamefont {Tartakovskii}}]{chekhovich15}%
  \BibitemOpen
  \bibfield  {author} {\bibinfo {author} {\bibfnamefont {E.~A.}\ \bibnamefont
  {Chekhovich}}, \bibinfo {author} {\bibfnamefont {M.}~\bibnamefont
  {Hopkinson}}, \bibinfo {author} {\bibfnamefont {M.~S.}\ \bibnamefont
  {Skolnick}}, \ and\ \bibinfo {author} {\bibfnamefont {A.~I.}\ \bibnamefont
  {Tartakovskii}},\ }\bibfield  {title} {\enquote {\bibinfo {title}
  {Suppression of nuclear spin bath fluctuations in self-assembled quantum dots
  induced by inhomogeneous strain},}\ }\href {\doibase 10.1038/ncomms7348}
  {\bibfield  {journal} {\bibinfo  {journal} {Nat. Commun.}\ }\textbf {\bibinfo
  {volume} {6}},\ \bibinfo {pages} {6348} (\bibinfo {year} {2015})}\BibitemShut
  {NoStop}%
\bibitem [{\citenamefont {Waeber}\ \emph {et~al.}(2016)\citenamefont {Waeber},
  \citenamefont {Hopkinson}, \citenamefont {Farrer}, \citenamefont {Ritchie},
  \citenamefont {Nilsson}, \citenamefont {Stevenson}, \citenamefont {Bennett},
  \citenamefont {Shields}, \citenamefont {Burkard}, \citenamefont
  {Tartakovskii}, \citenamefont {Skolnick},\ and\ \citenamefont
  {Chekhovich}}]{waeber16}%
  \BibitemOpen
  \bibfield  {author} {\bibinfo {author} {\bibfnamefont {A.~M.}\ \bibnamefont
  {Waeber}}, \bibinfo {author} {\bibfnamefont {M.}~\bibnamefont {Hopkinson}},
  \bibinfo {author} {\bibfnamefont {I.}~\bibnamefont {Farrer}}, \bibinfo
  {author} {\bibfnamefont {D.~A.}\ \bibnamefont {Ritchie}}, \bibinfo {author}
  {\bibfnamefont {J.}~\bibnamefont {Nilsson}}, \bibinfo {author} {\bibfnamefont
  {R.~Ma.}\ \bibnamefont {Stevenson}}, \bibinfo {author} {\bibfnamefont
  {A.~J.}\ \bibnamefont {Bennett}}, \bibinfo {author} {\bibfnamefont {A.~J.}\
  \bibnamefont {Shields}}, \bibinfo {author} {\bibfnamefont {G.}~\bibnamefont
  {Burkard}}, \bibinfo {author} {\bibfnamefont {A.~I.}\ \bibnamefont
  {Tartakovskii}}, \bibinfo {author} {\bibfnamefont {M.~S.}\ \bibnamefont
  {Skolnick}}, \ and\ \bibinfo {author} {\bibfnamefont {E.~A.}\ \bibnamefont
  {Chekhovich}},\ }\bibfield  {title} {\enquote {\bibinfo {title}
  {Few-second-long correlation times in a quantum dot nuclear spin bath probed
  by frequency-comb nuclear magnetic resonance spectroscopy},}\ }\href
  {\doibase 10.1038/nphys3686} {\bibfield  {journal} {\bibinfo  {journal} {Nat.
  Phys.}\ }\textbf {\bibinfo {volume} {12}},\ \bibinfo {pages} {688} (\bibinfo
  {year} {2016})}\BibitemShut {NoStop}%
\bibitem [{\citenamefont {Kuznetsova}\ \emph {et~al.}(2014)\citenamefont
  {Kuznetsova}, \citenamefont {Flisinski}, \citenamefont {Gerlovin},
  \citenamefont {Petrov}, \citenamefont {Ignatiev}, \citenamefont {Verbin},
  \citenamefont {Yakovlev}, \citenamefont {Reuter}, \citenamefont {Wieck},\
  and\ \citenamefont {Bayer}}]{kuznetsova14}%
  \BibitemOpen
  \bibfield  {author} {\bibinfo {author} {\bibfnamefont {M.~S.}\ \bibnamefont
  {Kuznetsova}}, \bibinfo {author} {\bibfnamefont {K.}~\bibnamefont
  {Flisinski}}, \bibinfo {author} {\bibfnamefont {I.~Ya.}\ \bibnamefont
  {Gerlovin}}, \bibinfo {author} {\bibfnamefont {M.~Yu.}\ \bibnamefont
  {Petrov}}, \bibinfo {author} {\bibfnamefont {I.~V.}\ \bibnamefont
  {Ignatiev}}, \bibinfo {author} {\bibfnamefont {S.~Yu.}\ \bibnamefont
  {Verbin}}, \bibinfo {author} {\bibfnamefont {D.~R.}\ \bibnamefont
  {Yakovlev}}, \bibinfo {author} {\bibfnamefont {D.}~\bibnamefont {Reuter}},
  \bibinfo {author} {\bibfnamefont {A.~D.}\ \bibnamefont {Wieck}}, \ and\
  \bibinfo {author} {\bibfnamefont {M.}~\bibnamefont {Bayer}},\ }\bibfield
  {title} {\enquote {\bibinfo {title} {Nuclear magnetic resonances in
  (in,ga)as/gaas quantum dots studied by resonant optical pumping},}\ }\href
  {\doibase 10.1103/PhysRevB.89.125304} {\bibfield  {journal} {\bibinfo
  {journal} {Phys. Rev. B}\ }\textbf {\bibinfo {volume} {89}},\ \bibinfo
  {pages} {125304} (\bibinfo {year} {2014})}\BibitemShut {NoStop}%
\bibitem [{\citenamefont {Munsch}\ \emph {et~al.}(2014)\citenamefont {Munsch},
  \citenamefont {W{\"u}st}, \citenamefont {Kuhlmann}, \citenamefont {Xue},
  \citenamefont {Ludwig}, \citenamefont {Reuter}, \citenamefont {Wieck},
  \citenamefont {Poggio},\ and\ \citenamefont {Warburton}}]{munsch14}%
  \BibitemOpen
  \bibfield  {author} {\bibinfo {author} {\bibfnamefont {M.}~\bibnamefont
  {Munsch}}, \bibinfo {author} {\bibfnamefont {G.}~\bibnamefont {W{\"u}st}},
  \bibinfo {author} {\bibfnamefont {A.~V.}\ \bibnamefont {Kuhlmann}}, \bibinfo
  {author} {\bibfnamefont {F.}~\bibnamefont {Xue}}, \bibinfo {author}
  {\bibfnamefont {A.}~\bibnamefont {Ludwig}}, \bibinfo {author} {\bibfnamefont
  {D.}~\bibnamefont {Reuter}}, \bibinfo {author} {\bibfnamefont {A.~D.}\
  \bibnamefont {Wieck}}, \bibinfo {author} {\bibfnamefont {M.}~\bibnamefont
  {Poggio}}, \ and\ \bibinfo {author} {\bibfnamefont {R.~J.}\ \bibnamefont
  {Warburton}},\ }\bibfield  {title} {\enquote {\bibinfo {title} {Manipulation
  of the nuclear spin ensemble in a quantum dot with chirped magnetic resonance
  pulses},}\ }\href {\doibase 10.1038/nnano.2014.175} {\bibfield  {journal}
  {\bibinfo  {journal} {Nat. Nanotechnol.}\ }\textbf {\bibinfo {volume} {9}},\
  \bibinfo {pages} {671--675} (\bibinfo {year} {2014})}\BibitemShut {NoStop}%
\bibitem [{\citenamefont {Ajoy}\ and\ \citenamefont
  {Cappellaro}(2012)}]{ajoy12}%
  \BibitemOpen
  \bibfield  {author} {\bibinfo {author} {\bibfnamefont {A.}~\bibnamefont
  {Ajoy}}\ and\ \bibinfo {author} {\bibfnamefont {P.}~\bibnamefont
  {Cappellaro}},\ }\bibfield  {title} {\enquote {\bibinfo {title} {Stable
  three-axis nuclear-spin gyroscope in diamond},}\ }\href {\doibase
  10.1103/PhysRevA.86.062104} {\bibfield  {journal} {\bibinfo  {journal} {Phys.
  Rev. A}\ }\textbf {\bibinfo {volume} {86}},\ \bibinfo {pages} {062104}
  (\bibinfo {year} {2012})}\BibitemShut {NoStop}%
\bibitem [{\citenamefont {He}\ \emph {et~al.}(1993)\citenamefont {He},
  \citenamefont {Manson},\ and\ \citenamefont {Fisk}}]{he93}%
  \BibitemOpen
  \bibfield  {author} {\bibinfo {author} {\bibfnamefont {X.-F.}\ \bibnamefont
  {He}}, \bibinfo {author} {\bibfnamefont {N.~B.}\ \bibnamefont {Manson}}, \
  and\ \bibinfo {author} {\bibfnamefont {P.~T.~H.}\ \bibnamefont {Fisk}},\
  }\bibfield  {title} {\enquote {\bibinfo {title} {Paramagnetic resonance of
  photoexcited n- \textit{V} defects in diamond. ii. hyperfine interaction with
  the $^{14}\mathrm{N}$ nucleus},}\ }\href {\doibase 10.1103/PhysRevB.47.8816}
  {\bibfield  {journal} {\bibinfo  {journal} {Phys. Rev. B}\ }\textbf {\bibinfo
  {volume} {47}},\ \bibinfo {pages} {8816--8822} (\bibinfo {year}
  {1993})}\BibitemShut {NoStop}%
\bibitem [{\citenamefont {Teles}\ \emph {et~al.}(2015)\citenamefont {Teles},
  \citenamefont {Rivera-Ascona}, \citenamefont {Polli}, \citenamefont
  {Oliveira-Silva}, \citenamefont {Vidoto}, \citenamefont {Andreeta},\ and\
  \citenamefont {Bonagamba}}]{teles15}%
  \BibitemOpen
  \bibfield  {author} {\bibinfo {author} {\bibfnamefont {J.}~\bibnamefont
  {Teles}}, \bibinfo {author} {\bibfnamefont {C.}~\bibnamefont
  {Rivera-Ascona}}, \bibinfo {author} {\bibfnamefont {R.~S.}\ \bibnamefont
  {Polli}}, \bibinfo {author} {\bibfnamefont {R.}~\bibnamefont
  {Oliveira-Silva}}, \bibinfo {author} {\bibfnamefont {E.~L.~G.}\ \bibnamefont
  {Vidoto}}, \bibinfo {author} {\bibfnamefont {J.~P.}\ \bibnamefont
  {Andreeta}}, \ and\ \bibinfo {author} {\bibfnamefont {T.~J.}\ \bibnamefont
  {Bonagamba}},\ }\bibfield  {title} {\enquote {\bibinfo {title} {Experimental
  implementation of quantum information processing by zeeman-perturbed nuclear
  quadrupole resonance},}\ }\href {\doibase 10.1007/s11128-015-0967-3}
  {\bibfield  {journal} {\bibinfo  {journal} {Quantum Inf. Process.}\ }\textbf
  {\bibinfo {volume} {14}},\ \bibinfo {pages} {1889--1906} (\bibinfo {year}
  {2015})}\BibitemShut {NoStop}%
\bibitem [{\citenamefont {Joana}\ \emph {et~al.}(2016)\citenamefont {Joana},
  \citenamefont {van Loock}, \citenamefont {Deng},\ and\ \citenamefont
  {Byrnes}}]{byrnes16}%
  \BibitemOpen
  \bibfield  {author} {\bibinfo {author} {\bibfnamefont {C.}~\bibnamefont
  {Joana}}, \bibinfo {author} {\bibfnamefont {P.}~\bibnamefont {van Loock}},
  \bibinfo {author} {\bibfnamefont {H.}~\bibnamefont {Deng}}, \ and\ \bibinfo
  {author} {\bibfnamefont {T.}~\bibnamefont {Byrnes}},\ }\bibfield  {title}
  {\enquote {\bibinfo {title} {Steady-state generation of
  negative-wigner-function light using feedback},}\ }\href {\doibase
  10.1103/PhysRevA.94.063802} {\bibfield  {journal} {\bibinfo  {journal} {Phys.
  Rev. A}\ }\textbf {\bibinfo {volume} {94}},\ \bibinfo {pages} {063802}
  (\bibinfo {year} {2016})}\BibitemShut {NoStop}%
\bibitem [{\citenamefont {Suter}\ and\ \citenamefont
  {\'Alvarez}(2016)}]{suter16}%
  \BibitemOpen
  \bibfield  {author} {\bibinfo {author} {\bibfnamefont {D.}~\bibnamefont
  {Suter}}\ and\ \bibinfo {author} {\bibfnamefont {G.~A.}\ \bibnamefont
  {\'Alvarez}},\ }\bibfield  {title} {\enquote {\bibinfo {title}
  {\textit{Colloquium} : Protecting quantum information against environmental
  noise},}\ }\href {\doibase 10.1103/RevModPhys.88.041001} {\bibfield
  {journal} {\bibinfo  {journal} {Rev. Mod. Phys.}\ }\textbf {\bibinfo {volume}
  {88}},\ \bibinfo {pages} {041001} (\bibinfo {year} {2016})}\BibitemShut
  {NoStop}%
\bibitem [{\citenamefont {Byrnes}\ \emph {et~al.}(2015)\citenamefont {Byrnes},
  \citenamefont {Rosseau}, \citenamefont {Khosla}, \citenamefont {Pyrkov},
  \citenamefont {Thomasen}, \citenamefont {Mukai}, \citenamefont {Koyama},
  \citenamefont {Abdelrahman},\ and\ \citenamefont {Ilo-Okeke}}]{byrnes15}%
  \BibitemOpen
  \bibfield  {author} {\bibinfo {author} {\bibfnamefont {T.}~\bibnamefont
  {Byrnes}}, \bibinfo {author} {\bibfnamefont {D.}~\bibnamefont {Rosseau}},
  \bibinfo {author} {\bibfnamefont {M.}~\bibnamefont {Khosla}}, \bibinfo
  {author} {\bibfnamefont {A.}~\bibnamefont {Pyrkov}}, \bibinfo {author}
  {\bibfnamefont {A.}~\bibnamefont {Thomasen}}, \bibinfo {author}
  {\bibfnamefont {T.}~\bibnamefont {Mukai}}, \bibinfo {author} {\bibfnamefont
  {S.}~\bibnamefont {Koyama}}, \bibinfo {author} {\bibfnamefont
  {A.}~\bibnamefont {Abdelrahman}}, \ and\ \bibinfo {author} {\bibfnamefont
  {E.}~\bibnamefont {Ilo-Okeke}},\ }\bibfield  {title} {\enquote {\bibinfo
  {title} {Macroscopic quantum information processing using spin coherent
  states},}\ }\href {\doibase http://dx.doi.org/10.1016/j.optcom.2014.08.017}
  {\bibfield  {journal} {\bibinfo  {journal} {Optics Commun.}\ }\textbf
  {\bibinfo {volume} {337}},\ \bibinfo {pages} {102 -- 109} (\bibinfo {year}
  {2015})}\BibitemShut {NoStop}%
\bibitem [{\citenamefont {Waldherr}\ and\ \citenamefont
  {et~al.}(2014)}]{waldherr14}%
  \BibitemOpen
  \bibfield  {author} {\bibinfo {author} {\bibfnamefont {G.}~\bibnamefont
  {Waldherr}}\ and\ \bibinfo {author} {\bibnamefont {et~al.}},\ }\bibfield
  {title} {\enquote {\bibinfo {title} {Quantum error correction in a
  solid-state hybrid spin register},}\ }\href@noop {} {\bibfield  {journal}
  {\bibinfo  {journal} {Nature (London)}\ }\textbf {\bibinfo {volume} {506}},\
  \bibinfo {pages} {204--207} (\bibinfo {year} {2014})}\BibitemShut {NoStop}%
\bibitem [{\citenamefont {Manson}\ \emph {et~al.}(2006)\citenamefont {Manson},
  \citenamefont {Harrison},\ and\ \citenamefont {Sellars}}]{manson06}%
  \BibitemOpen
  \bibfield  {author} {\bibinfo {author} {\bibfnamefont {N.~B.}\ \bibnamefont
  {Manson}}, \bibinfo {author} {\bibfnamefont {J.~P.}\ \bibnamefont
  {Harrison}}, \ and\ \bibinfo {author} {\bibfnamefont {M.~J.}\ \bibnamefont
  {Sellars}},\ }\bibfield  {title} {\enquote {\bibinfo {title}
  {Nitrogen-vacancy center in diamond: Model of the electronic structure and
  associated dynamics},}\ }\href {\doibase 10.1103/PhysRevB.74.104303}
  {\bibfield  {journal} {\bibinfo  {journal} {Phys. Rev. B}\ }\textbf {\bibinfo
  {volume} {74}},\ \bibinfo {pages} {104303} (\bibinfo {year}
  {2006})}\BibitemShut {NoStop}%
\bibitem [{\citenamefont {Smeltzer}\ \emph {et~al.}(2009)\citenamefont
  {Smeltzer}, \citenamefont {McIntyre},\ and\ \citenamefont
  {Childress}}]{smeltzer09}%
  \BibitemOpen
  \bibfield  {author} {\bibinfo {author} {\bibfnamefont {B.}~\bibnamefont
  {Smeltzer}}, \bibinfo {author} {\bibfnamefont {J.}~\bibnamefont {McIntyre}},
  \ and\ \bibinfo {author} {\bibfnamefont {L.}~\bibnamefont {Childress}},\
  }\bibfield  {title} {\enquote {\bibinfo {title} {Robust control of individual
  nuclear spins in diamond},}\ }\href {\doibase 10.1103/PhysRevA.80.050302}
  {\bibfield  {journal} {\bibinfo  {journal} {Phys. Rev. A}\ }\textbf {\bibinfo
  {volume} {80}},\ \bibinfo {pages} {050302} (\bibinfo {year}
  {2009})}\BibitemShut {NoStop}%
\bibitem [{\citenamefont {Morton}\ \emph {et~al.}(2006)\citenamefont {Morton},
  \citenamefont {Tyryshkin}, \citenamefont {Ardavan}, \citenamefont {Benjamin},
  \citenamefont {Porfyrakis}, \citenamefont {Lyon},\ and\ \citenamefont
  {Briggs}}]{morton06}%
  \BibitemOpen
  \bibfield  {author} {\bibinfo {author} {\bibfnamefont {J.~J.~L.}\
  \bibnamefont {Morton}}, \bibinfo {author} {\bibfnamefont {A.~M.}\
  \bibnamefont {Tyryshkin}}, \bibinfo {author} {\bibfnamefont {A.}~\bibnamefont
  {Ardavan}}, \bibinfo {author} {\bibfnamefont {S.~C.}\ \bibnamefont
  {Benjamin}}, \bibinfo {author} {\bibfnamefont {K.}~\bibnamefont
  {Porfyrakis}}, \bibinfo {author} {\bibfnamefont {S.~A.}\ \bibnamefont
  {Lyon}}, \ and\ \bibinfo {author} {\bibfnamefont {G.~A.~D.}\ \bibnamefont
  {Briggs}},\ }\bibfield  {title} {\enquote {\bibinfo {title} {Bang--bang
  control of fullerene qubits using ultrafast phase gates},}\ }\href@noop {}
  {\bibfield  {journal} {\bibinfo  {journal} {Nat. Phys.}\ }\textbf {\bibinfo
  {volume} {2}},\ \bibinfo {pages} {40--43} (\bibinfo {year}
  {2006})}\BibitemShut {NoStop}%
\bibitem [{\citenamefont {Sangtawesin}\ \emph {et~al.}(2014)\citenamefont
  {Sangtawesin}, \citenamefont {Brundage},\ and\ \citenamefont
  {Petta}}]{sangtawesin14}%
  \BibitemOpen
  \bibfield  {author} {\bibinfo {author} {\bibfnamefont {S.}~\bibnamefont
  {Sangtawesin}}, \bibinfo {author} {\bibfnamefont {T.~O.}\ \bibnamefont
  {Brundage}}, \ and\ \bibinfo {author} {\bibfnamefont {J.~R.}\ \bibnamefont
  {Petta}},\ }\bibfield  {title} {\enquote {\bibinfo {title} {Fast
  room-temperature phase gate on a single nuclear spin in diamond},}\ }\href
  {\doibase 10.1103/PhysRevLett.113.020506} {\bibfield  {journal} {\bibinfo
  {journal} {Phys. Rev. Lett.}\ }\textbf {\bibinfo {volume} {113}},\ \bibinfo
  {pages} {020506} (\bibinfo {year} {2014})}\BibitemShut {NoStop}%
\bibitem [{\citenamefont {Chuang}\ \emph {et~al.}(1998)\citenamefont {Chuang},
  \citenamefont {Gershenfeld}, \citenamefont {Kubinec},\ and\ \citenamefont
  {Leung}}]{chuang98}%
  \BibitemOpen
  \bibfield  {author} {\bibinfo {author} {\bibfnamefont {I.~L.}\ \bibnamefont
  {Chuang}}, \bibinfo {author} {\bibfnamefont {N.}~\bibnamefont {Gershenfeld}},
  \bibinfo {author} {\bibfnamefont {M.~G.}\ \bibnamefont {Kubinec}}, \ and\
  \bibinfo {author} {\bibfnamefont {D.~W.}\ \bibnamefont {Leung}},\ }\bibfield
  {title} {\enquote {\bibinfo {title} {Bulk quantum computation with nuclear
  magnetic resonance: theory and experiment},}\ }\href {\doibase
  10.1098/rspa.1998.0170} {\bibfield  {journal} {\bibinfo  {journal} {Proc. R.
  Soc. London, Ser. A}\ }\textbf {\bibinfo {volume} {454}},\ \bibinfo {pages}
  {447--467} (\bibinfo {year} {1998})}\BibitemShut {NoStop}%
\bibitem [{\citenamefont {Cramer}\ \emph {et~al.}(2016)\citenamefont {Cramer},
  \citenamefont {Kalb}, \citenamefont {Rol}, \citenamefont {Hensen},
  \citenamefont {Blok}, \citenamefont {Markham}, \citenamefont {Twitchen},
  \citenamefont {Hanson},\ and\ \citenamefont {Taminiau}}]{cramer16}%
  \BibitemOpen
  \bibfield  {author} {\bibinfo {author} {\bibfnamefont {J.}~\bibnamefont
  {Cramer}}, \bibinfo {author} {\bibfnamefont {N.}~\bibnamefont {Kalb}},
  \bibinfo {author} {\bibfnamefont {M.~A.}\ \bibnamefont {Rol}}, \bibinfo
  {author} {\bibfnamefont {B.}~\bibnamefont {Hensen}}, \bibinfo {author}
  {\bibfnamefont {M.~S.}\ \bibnamefont {Blok}}, \bibinfo {author}
  {\bibfnamefont {M.}~\bibnamefont {Markham}}, \bibinfo {author} {\bibfnamefont
  {D.~J.}\ \bibnamefont {Twitchen}}, \bibinfo {author} {\bibfnamefont
  {R.}~\bibnamefont {Hanson}}, \ and\ \bibinfo {author} {\bibfnamefont {T.~H.}\
  \bibnamefont {Taminiau}},\ }\bibfield  {title} {\enquote {\bibinfo {title}
  {Repeated quantum error correction on a continuously encoded qubit by
  real-time feedback},}\ }\href {\doibase 10.1038/ncomms11526} {\bibfield
  {journal} {\bibinfo  {journal} {Nat. Commun.}\ }\textbf {\bibinfo {volume}
  {7}},\ \bibinfo {pages} {11526} (\bibinfo {year} {2016})}\BibitemShut
  {NoStop}%
\bibitem [{\citenamefont {Teles}\ \emph {et~al.}(2007)\citenamefont {Teles},
  \citenamefont {deAzevedo}, \citenamefont {Auccaise}, \citenamefont
  {Sarthour}, \citenamefont {Oliveira},\ and\ \citenamefont
  {Bonagamba}}]{teles07}%
  \BibitemOpen
  \bibfield  {author} {\bibinfo {author} {\bibfnamefont {J.}~\bibnamefont
  {Teles}}, \bibinfo {author} {\bibfnamefont {E.~R.}\ \bibnamefont
  {deAzevedo}}, \bibinfo {author} {\bibfnamefont {R.}~\bibnamefont {Auccaise}},
  \bibinfo {author} {\bibfnamefont {R.~S.}\ \bibnamefont {Sarthour}}, \bibinfo
  {author} {\bibfnamefont {I.~S.}\ \bibnamefont {Oliveira}}, \ and\ \bibinfo
  {author} {\bibfnamefont {T.~J.}\ \bibnamefont {Bonagamba}},\ }\bibfield
  {title} {\enquote {\bibinfo {title} {Quantum state tomography for quadrupolar
  nuclei using global rotations of the spin system},}\ }\href {\doibase
  10.1063/1.2717179} {\bibfield  {journal} {\bibinfo  {journal} {J. Chem.
  Phys.}\ }\textbf {\bibinfo {volume} {126}},\ \bibinfo {pages} {154506}
  (\bibinfo {year} {2007})}\BibitemShut {NoStop}%
\bibitem [{\citenamefont {Rundle}\ \emph {et~al.}(2016)\citenamefont {Rundle},
  \citenamefont {Tilma}, \citenamefont {Samson},\ and\ \citenamefont
  {Everitt}}]{rundle16}%
  \BibitemOpen
  \bibfield  {author} {\bibinfo {author} {\bibfnamefont {R.~P.}\ \bibnamefont
  {Rundle}}, \bibinfo {author} {\bibfnamefont {T.}~\bibnamefont {Tilma}},
  \bibinfo {author} {\bibfnamefont {J.~H.}\ \bibnamefont {Samson}}, \ and\
  \bibinfo {author} {\bibfnamefont {M.~J.}\ \bibnamefont {Everitt}},\
  }\bibfield  {title} {\enquote {\bibinfo {title} {Quantum state reconstruction
  made easy: a direct method for tomography},}\ }\href@noop {} {\bibfield
  {journal} {\bibinfo  {journal} {arXiv preprint arXiv:1605.08922}\ } (\bibinfo
  {year} {2016})}\BibitemShut {NoStop}%
\bibitem [{\citenamefont {Gedik}\ \emph {et~al.}(2015)\citenamefont {Gedik},
  \citenamefont {Silva}, \citenamefont {{\c{C}}akmak}, \citenamefont {Karpat},
  \citenamefont {Vidoto}, \citenamefont {Soares-Pinto}, \citenamefont
  {deAzevedo},\ and\ \citenamefont {Fanchini}}]{gedik15}%
  \BibitemOpen
  \bibfield  {author} {\bibinfo {author} {\bibfnamefont {Z.}~\bibnamefont
  {Gedik}}, \bibinfo {author} {\bibfnamefont {I.~A.}\ \bibnamefont {Silva}},
  \bibinfo {author} {\bibfnamefont {B.}~\bibnamefont {{\c{C}}akmak}}, \bibinfo
  {author} {\bibfnamefont {G.}~\bibnamefont {Karpat}}, \bibinfo {author}
  {\bibfnamefont {E.~L.~G.}\ \bibnamefont {Vidoto}}, \bibinfo {author}
  {\bibfnamefont {D.~O.}\ \bibnamefont {Soares-Pinto}}, \bibinfo {author}
  {\bibfnamefont {E.~R.}\ \bibnamefont {deAzevedo}}, \ and\ \bibinfo {author}
  {\bibfnamefont {F.~F.}\ \bibnamefont {Fanchini}},\ }\bibfield  {title}
  {\enquote {\bibinfo {title} {Computational speed-up with a single qudit},}\
  }\href {\doibase 10.1103/10.1038/srep14671} {\bibfield  {journal} {\bibinfo
  {journal} {Sci. Rep.}\ }\textbf {\bibinfo {volume} {5}},\ \bibinfo {pages} {14671}
  (\bibinfo {year}{2015})}\BibitemShut {NoStop}%
\bibitem [{\citenamefont {Steffen}\ \emph {et~al.}(2016)\citenamefont
  {Steffen}, \citenamefont {Gambetta},\ and\ \citenamefont {Chow}}]{steffen16}%
  \BibitemOpen
  \bibfield  {author} {\bibinfo {author} {\bibfnamefont {M.}~\bibnamefont
  {Steffen}}, \bibinfo {author} {\bibfnamefont {J.~M.}\ \bibnamefont
  {Gambetta}}, \ and\ \bibinfo {author} {\bibfnamefont {J.~M.}\ \bibnamefont
  {Chow}},\ }\bibfield  {title} {\enquote {\bibinfo {title} {Progress, status,
  and prospects of superconducting qubits for quantum computing},}\ }in\ \href
  {\doibase 10.1109/ESSDERC.2016.7599578} {\emph {\bibinfo {booktitle} {2016
  46th European Solid-State Device Research Conference (ESSDERC)}}}\ (\bibinfo
  {year} {IEEE, 2016})\ pp.\ \bibinfo {pages} {17--20}\BibitemShut {NoStop}%
\bibitem [{\citenamefont {Leuenberger}\ and\ \citenamefont
  {Loss}(2003)}]{leuenberger03}%
  \BibitemOpen
  \bibfield  {author} {\bibinfo {author} {\bibfnamefont {M.~N.}\ \bibnamefont
  {Leuenberger}}\ and\ \bibinfo {author} {\bibfnamefont {D.}~\bibnamefont
  {Loss}},\ }\bibfield  {title} {\enquote {\bibinfo {title} {Grover algorithm
  for large nuclear spins in semiconductors},}\ }\href {\doibase
  10.1103/PhysRevB.68.165317} {\bibfield  {journal} {\bibinfo  {journal} {Phys.
  Rev. B}\ }\textbf {\bibinfo {volume} {536}},\ \bibinfo {pages} {441}
  (\bibinfo {year} {2003})}\BibitemShut {NoStop}%
\bibitem [{\citenamefont {Ofek}\ and\ \citenamefont {et~al.}(2016)}]{ofek16}%
  \BibitemOpen
  \bibfield  {author} {\bibinfo {author} {\bibfnamefont {N.}~\bibnamefont
  {Ofek}}\ and\ \bibinfo {author} {\bibnamefont {et~al.}},\ }\bibfield  {title}
  {\enquote {\bibinfo {title} {Extending the lifetime of a quantum bit with
  error correction in superconducting circuits},}\ }\href {\doibase
  10.1038/nature18949} {\bibfield  {journal} {\bibinfo  {journal} {Nature
  (London)}\ }\textbf {\bibinfo {volume} {68}},\ \bibinfo {pages} {165317}
 (\bibinfo {year} {2016})}\BibitemShut
  {NoStop}%
\bibitem [{\citenamefont {Hausmann}\ \emph {et~al.}(2011)\citenamefont
  {Hausmann}, \citenamefont {Babinec}, \citenamefont {Choy}, \citenamefont
  {Hodges}, \citenamefont {Hong}, \citenamefont {Bulu}, \citenamefont {Yacoby},
  \citenamefont {Lukin},\ and\ \citenamefont {Lon{\v{c}}ar}}]{hausmann11}%
  \BibitemOpen
  \bibfield  {author} {\bibinfo {author} {\bibfnamefont {B.~J.~M.}\
  \bibnamefont {Hausmann}}, \bibinfo {author} {\bibfnamefont {T.~M.}\
  \bibnamefont {Babinec}}, \bibinfo {author} {\bibfnamefont {J.~T.}\
  \bibnamefont {Choy}}, \bibinfo {author} {\bibfnamefont {J.~S.}\ \bibnamefont
  {Hodges}}, \bibinfo {author} {\bibfnamefont {S.}~\bibnamefont {Hong}},
  \bibinfo {author} {\bibfnamefont {I.}~\bibnamefont {Bulu}}, \bibinfo {author}
  {\bibfnamefont {A.}~\bibnamefont {Yacoby}}, \bibinfo {author} {\bibfnamefont
  {M.~D.}\ \bibnamefont {Lukin}}, \ and\ \bibinfo {author} {\bibfnamefont
  {M.}~\bibnamefont {Lon{\v{c}}ar}},\ }\bibfield  {title} {\enquote {\bibinfo
  {title} {Single-color centers implanted in diamond nanostructures},}\
  }\href@noop {} {\bibfield  {journal} {\bibinfo  {journal} {New J. Phys.}\
  }\textbf {\bibinfo {volume} {13}},\ \bibinfo {pages} {045004} (\bibinfo
  {year} {2011})}\BibitemShut {NoStop}%
\bibitem [{\citenamefont {Yang}\ and\ \citenamefont {Liu}(2009)}]{yang09}%
  \BibitemOpen
  \bibfield  {author} {\bibinfo {author} {\bibfnamefont {W.}~\bibnamefont
  {Yang}}\ and\ \bibinfo {author} {\bibfnamefont {R.-B.}\ \bibnamefont {Liu}},\
  }\bibfield  {title} {\enquote {\bibinfo {title} {Quantum many-body theory of
  qubit decoherence in a finite-size spin bath. ii. ensemble dynamics},}\
  }\href {\doibase 10.1103/PhysRevB.79.115320} {\bibfield  {journal} {\bibinfo
  {journal} {Phys. Rev. B}\ }\textbf {\bibinfo {volume} {79}},\ \bibinfo
  {pages} {115320} (\bibinfo {year} {2009})}\BibitemShut {NoStop}%
\bibitem [{\citenamefont {Stanek}\ \emph {et~al.}(2014)\citenamefont {Stanek},
  \citenamefont {Raas},\ and\ \citenamefont {Uhrig}}]{stanek14}%
  \BibitemOpen
  \bibfield  {author} {\bibinfo {author} {\bibfnamefont {D.}~\bibnamefont
  {Stanek}}, \bibinfo {author} {\bibfnamefont {C.}~\bibnamefont {Raas}}, \ and\
  \bibinfo {author} {\bibfnamefont {G.~S.}\ \bibnamefont {Uhrig}},\ }\bibfield
  {title} {\enquote {\bibinfo {title} {From quantum-mechanical to classical
  dynamics in the central-spin model},}\ }\href {\doibase
  10.1103/PhysRevB.90.064301} {\bibfield  {journal} {\bibinfo  {journal} {Phys.
  Rev. B}\ }\textbf {\bibinfo {volume} {90}},\ \bibinfo {pages} {064301}
  (\bibinfo {year} {2014})}\BibitemShut {NoStop}%
\bibitem [{\citenamefont {Bulutay}\ \emph {et~al.}(2014)\citenamefont
  {Bulutay}, \citenamefont {Chekhovich},\ and\ \citenamefont
  {Tartakovskii}}]{bulutay14}%
  \BibitemOpen
  \bibfield  {author} {\bibinfo {author} {\bibfnamefont {C.}~\bibnamefont
  {Bulutay}}, \bibinfo {author} {\bibfnamefont {E.~A.}\ \bibnamefont
  {Chekhovich}}, \ and\ \bibinfo {author} {\bibfnamefont {A.~I.}\ \bibnamefont
  {Tartakovskii}},\ }\bibfield  {title} {\enquote {\bibinfo {title} {Nuclear
  magnetic resonance inverse spectra of ${\rm ingaas}$ quantum dots: Atomistic
  level structural information},}\ }\href {\doibase 10.1103/PhysRevB.90.205425}
  {\bibfield  {journal} {\bibinfo  {journal} {Phys. Rev. B}\ }\textbf {\bibinfo
  {volume} {90}},\ \bibinfo {pages} {205425} (\bibinfo {year}
  {2014})}\BibitemShut {NoStop}%
\bibitem [{\citenamefont {Knill}(2005)}]{knill05}%
  \BibitemOpen
  \bibfield  {author} {\bibinfo {author} {\bibfnamefont {E.}~\bibnamefont
  {Knill}},\ }\bibfield  {title} {\enquote {\bibinfo {title} {Quantum computing
  with realistically noisy devices},}\ }\href@noop {} {\bibfield  {journal}
  {\bibinfo  {journal} {Nature (London)}\ }\textbf {\bibinfo {volume} {434}},\
  \bibinfo {pages} {39--44} (\bibinfo {year} {2005})}\BibitemShut {NoStop}%
\bibitem [{\citenamefont {De~Zela}(2014)}]{dezela14}%
  \BibitemOpen
  \bibfield  {author} {\bibinfo {author} {\bibfnamefont {F.}~\bibnamefont
  {De~Zela}},\ }\bibfield  {title} {\enquote {\bibinfo {title} {Closed-form
  expressions for the matrix exponential},}\ }\href {\doibase
  10.3390/sym6020329} {\bibfield  {journal} {\bibinfo  {journal} {Symmetry}\
  }\textbf {\bibinfo {volume} {6}},\ \bibinfo {pages} {329--344} (\bibinfo
  {year} {2014})}\BibitemShut {NoStop}%
\end{thebibliography}
\end{document}